\documentstyle{article}
\def\emline#1#2#3#4#5#6{%
       \put(#1,#2){\special{em:moveto}}%
       \put(#4,#5){\special{em:lineto}}}
\def\newpic#1{}

\catcode`\@=11
\def\marginnote#1{}

\newcount\hour
\newcount\minute
\newtoks\amorpm
\hour=\time\divide\hour by60
\minute=\time{\multiply\hour by60 \global\advance\minute by-\hour}
\edef\standardtime{{\ifnum\hour<12 \global\amorpm={am}%
        \else\global\amorpm={pm}\advance\hour by-12 \fi
        \ifnum\hour=0 \hour=12 \fi
        \number\hour:\ifnum\minute<10 0\fi\number\minute\the\amorpm}}
\edef\militarytime{\number\hour:\ifnum\minute<10 0\fi\number\minute}

\def\draftlabel#1{{\@bsphack\if@filesw {\let\thepage\relax
   \xdef\@gtempa{\write\@auxout{\string
      \newlabel{#1}{{\@currentlabel}{\thepage}}}}}\@gtempa
   \if@nobreak \ifvmode\nobreak\fi\fi\fi\@esphack}
        \gdef\@eqnlabel{#1}}
\def\@eqnlabel{}
\def\@vacuum{}
\def\draftmarginnote#1{\marginpar{\raggedright\scriptsize\tt#1}}

\def\draft{\oddsidemargin -.5truein
        \def\@oddfoot{\sl preliminary draft \hfil
        \rm\thepage\hfil\sl\today\quad\militarytime}
        \let\@evenfoot\@oddfoot \overfullrule 3pt
        \let\label=\draftlabel
        \let\marginnote=\draftmarginnote
   \def\@eqnnum{(\theequation)\rlap{\kern\marginparsep\tt\@eqnlabel}%
\global\let\@eqnlabel\@vacuum}  }
\begingroup\ifx\undefined\newsymbol \else\def\input#1 {\endgroup}\fi
\input amssym.def \relax
\input amssym
\newfont{\hr}{msbm10}
\newfont{\ams}{msam10}
\def\RS{Ruijsenaars-Schneider\ }
\def\XXX{{\bf XXX\ }}
\def\XXZ{{\bf XXZ\ }}
\def\XYZ{{\bf XYZ\ }}

\def \tr{{\rm tr}}

\def \det{{\rm det}}

\def \log {{\rm log}}

\def \sn{{\rm sn}}
\def \cn{{\rm cn}}
\def \dn{{\rm dn}}

\def\tilde{\widetilde}
\def\bar{\overline}
\def\hat{\widehat}
\def\*{\star}
\def\({\left(}		
\def\){\right)}		
\def\[{\left[}		
\def\]{\right]}

\def\frac#1#2{{#1 \over #2}}

\def\2pi{\hbox{$2\pi i$}}

\def\dsl{\raise.15ex\hbox{/}\kern-.57em\partial}
\def\Dsl{\,\raise.15ex\hbox{/}\mkern-.13.5mu D}
		\def\CF{{\cal F}}

	\def\CN{{\cal N}}

\font\numbers=cmss12
\font\upright=cmu10 scaled\magstep1
\def\stroke{\vrule height8pt width0.4pt depth-0.1pt}
\def\topfleck{\vrule height8pt width0.5pt depth-5.9pt}
\def\botfleck{\vrule height2pt width0.5pt depth0.1pt}
\def\Zmath{\vcenter{\hbox{\numbers\rlap{\rlap{Z}\kern 0.8pt\topfleck}\kern
2.2pt
                   \rlap Z\kern 6pt\botfleck\kern 1pt}}}
\def\Qmath{\vcenter{\hbox{\upright\rlap{\rlap{Q}\kern
                   3.8pt\stroke}\phantom{Q}}}}
\def\Nmath{\vcenter{\hbox{\upright\rlap{I}\kern 1.7pt N}}}
\def\Cmath{\vcenter{\hbox{\upright\rlap{\rlap{C}\kern
                   3.8pt\stroke}\phantom{C}}}}
\def\Rmath{\vcenter{\hbox{\upright\rlap{I}\kern 1.7pt R}}}
\def\Z{\ifmmode\Zmath\else$\Zmath$\fi}
\def\Q{\ifmmode\Qmath\else$\Qmath$\fi}
\def\N{\ifmmode\Nmath\else$\Nmath$\fi}
\def\C{\ifmmode\Cmath\else$\Cmath$\fi}
\def\R{\ifmmode\Rmath\else$\Rmath$\fi}
\newcounter{app}

\def\app{\setcounter{equation}{0}
\def\theequation{\Alph{app}.\arabic{equation}}\par
   \addvspace{4ex}
   \@afterindentfalse
  \secdef\@app\@dapp}

\newcommand\@app{\@startsection {app}{1}{0ex}%
                                   {-3.5ex \@plus -1ex \@minus -.2ex}%
                                   {2.3ex \@plus.2ex}%
                                   {\normalfont\Large\bf}}
\def\@dapp#1{%
{\parindent \z@ \raggedright  \bf #1}\par\nobreak}
\def\l@app#1#2{\ifnum \c@tocdepth >\z@
    \addpenalty\@secpenalty
    \addvspace{1.0em \@plus\p@}%
    \setlength\@tempdima{8em}%
    \begingroup
      \parindent \z@ \rightskip \@pnumwidth
      \parfillskip -\@pnumwidth
      \leavevmode \bfseries
      \advance\leftskip\@tempdima
      \hskip -\leftskip
      #1\nobreak\hfil \nobreak\hb@xt@\@pnumwidth{\hss #2}\par
    \endgroup\fi}
\def\stackreb#1#2{\mathrel{\mathop{#2}\limits_{#1}}}
\def\Tr{{\rm Tr}}
\def\res{{\rm res}}
\def\Bf#1{\mbox{\boldmath $#1$}}

\def\bphi{{\Bf\phi}}

\def\Im{{\rm Im}}

\def\2{{1\over 2}}
\def\N2{${\cal N}=2$}
\def\n=1{$\CN=1$}

\makeatletter
\newdimen\normalarrayskip              % skip between lines
\newdimen\minarrayskip                 % minimal skip between lines
\normalarrayskip\baselineskip
\minarrayskip\jot
\newif\ifold             \oldtrue            \def\new{\oldfalse}
\def\arraymode{\ifold\relax\else\displaystyle\fi} % mode of array entries
\def\eqnumphantom{\phantom{(\theequation)}}     % right phantom in eqnarray
\def\@arrayskip{\ifold\baselineskip\z@\lineskip\z@
     \else
     \baselineskip\minarrayskip\lineskip2\minarrayskip\fi}
\def\@arrayclassz{\ifcase \@lastchclass \@acolampacol \or
\@ampacol \or \or \or \@addamp \or
   \@acolampacol \or \@firstampfalse \@acol \fi
\edef\@preamble{\@preamble
  \ifcase \@chnum
     \hfil$\relax\arraymode\@sharp$\hfil
     \or $\relax\arraymode\@sharp$\hfil
     \or \hfil$\relax\arraymode\@sharp$\fi}}
\def\@array[#1]#2{\setbox\@arstrutbox=\hbox{\vrule
     height\arraystretch \ht\strutbox
     depth\arraystretch \dp\strutbox
     width\z@}\@mkpream{#2}\edef\@preamble{\halign
\noexpand\@halignto
\bgroup \tabskip\z@ \@arstrut \@preamble \tabskip\z@ \cr}%
\let\@startpbox\@@startpbox \let\@endpbox\@@endpbox
  \if #1t\vtop \else \if#1b\vbox \else \vcenter \fi\fi
  \bgroup \let\par\relax
  \let\@sharp##\let\protect\relax
  \@arrayskip\@preamble}
\def\eqnarray{\stepcounter{equation}%
              \let\@currentlabel=\theequation
              \global\@eqnswtrue
              \global\@eqcnt\z@
              \tabskip\@centering
              \let\\=\@eqncr
              $$%
 \halign to \displaywidth\bgroup
    \eqnumphantom\@eqnsel\hskip\@centering
    $\displaystyle \tabskip\z@ {##}$%
    \global\@eqcnt\@ne \hskip 2\arraycolsep
         $\displaystyle\arraymode{##}$\hfil
    \global\@eqcnt\tw@ \hskip 2\arraycolsep
         $\displaystyle\tabskip\z@{##}$\hfil
         \tabskip\@centering
    &{##}\tabskip\z@\cr}
\def\theequation{\thesection.\arabic{equation}}

\def\IB{\relax\hbox{$\inbar\kern-.3em{\rm B}$}}
\def\IC{\relax\hbox{$\inbar\kern-.3em{\rm C}$}}
\def\ID{\relax\hbox{$\inbar\kern-.3em{\rm D}$}}
\def\IE{\relax\hbox{$\inbar\kern-.3em{\rm E}$}}
\def\IF{\relax\hbox{$\inbar\kern-.3em{\rm F}$}}
\def\IG{\relax\hbox{$\inbar\kern-.3em{\rm G}$}}
\def\IGa{\relax\hbox{${\rm I}\kern-.18em\Gamma$}}
\def\IH{\relax{\rm I\kern-.18em H}}
\def\IK{\relax{\rm I\kern-.18em K}}
\def\IL{\relax{\rm I\kern-.18em L}}
\def\IP{\relax{\rm I\kern-.18em P}}
\def\IR{\relax{\rm I\kern-.18em R}}
\def\IZ{\relax\ifmmode\mathchoice
{\hbox{\cmss Z\kern-.4em Z}}{\hbox{\cmss Z\kern-.4em Z}}
{\lower.9pt\hbox{\cmsss Z\kern-.4em Z}}
{\lower1.2pt\hbox{\cmsss Z\kern-.4em Z}}\else{\cmss Z\kern-.4em Z}\fi}
\def\ocom{\relax{\raise1.2pt\hbox{$\otimes$}\kern-.5em\lower.1pt\hbox{,}}\,}
\font\manual=manfnt \def\dbend{\lower3.5pt\hbox{\manual\char127}}

\def\inbar{\,\vrule height1.5ex width.4pt depth0pt}
\def\bea{\begin{eqnarray}}
\def\eea{\end{eqnarray}}
\def\nn{\nonumber}
\def\beq{\begin{equation}}
\def\eeq{\end{equation}}
\def\ba{\beq\new\begin{array}{c}}
\def\ea{\end{array}\eeq}
\def\be{\ba}
\def\ee{\ea}
\def\stackreb#1#2{\mathrel{\mathop{#2}\limits_{#1}}}
\def\f{1\over}
\def\SW{Seiberg-Witten theory\ }

\def\Q{\bar{Q}}

\begin{document}
\begin{titlepage}
\setcounter{footnote}0
\def\thefootnote{\fnsymbol{footnote}}
\begin{center}
\hfill FIAN/TD-29/00\\
\hfill ITEP/TH-63/00\\
\hfill hep-th/0011093\\
\vspace{0.4in}
{\LARGE\bf Seiberg-Witten theory and duality in integrable systems}\\
\bigskip
\bigskip
\bigskip
{\Large A.Mironov
\footnote{E-mail:
mironov@lpi.ru, mironov@itep.ru}}\\
\bigskip 
{Theory Department, Lebedev Physics 
Institute, Moscow ~117924, Russia\\ 
and ITEP, Moscow 117259, Russia}
\end{center}
\bigskip \bigskip

\begin{abstract}
These lectures are devoted to the low energy limit of \N2 SUSY gauge 
theories, which is described in terms of integrable systems. A special 
emphasis is on a duality that naturally acts on these integrable systems. 
The duality turns out to be an effective tool in constructing the double 
elliptic integrable system which describes the six-dimensional 
Seiberg-Witten theory. At the same time, it implies a series of relations 
between other Seiberg-Witten systems. 
\end{abstract}

\end{titlepage}

\part{Seiberg-Witten theory: gauge versus integrable theories}
\setcounter{equation}{0}
\setcounter{footnote}{0}
\def\thefootnote{\arabic{footnote}}

The main target of these lectures is to present an integrable framework for one
of the most intriguing discoveries of recent years in quantum field theory --
a
possibility of exact, non-perturbative treating the low energy behaviour 
of supersymmetric gauge theories proposed by N.Seiberg and E.Witten 
\cite{SW1,SW2}. The results obtained turned out
to be most effectively described by integrable systems \cite{GKMMM}. Since
then,
many new features of Seiberg-Witten theories have been discovered, all of them
this or that way related to integrable systems (an intensive
discussion of integrability in \SW and related structures (Whitham hierarchies
and WDVV equations) can be found in the book \cite{BK} that contains 
recent comprehensive reviews).  Here we are going to concentrate on the most 
recent and puzzling structure, a duality present in integrable systems
of the Toda/Calogero/Ruijsenaars family \cite{rud}.

Therefore, technically the lectures are divided into two parts. The first part
is devoted to the Seiberg-Witten theories and their treatment within the
integrable approach. The second part deals with the duality and its role in
the
Seiberg-Witten theories.

\section{Seiberg-Witten solution}

Being one of the most fascinating breakthroughs in field theory, the 
Seiberg-Witten approach 
opened new possibilities to deal with nonperturbative physics at
strong coupling that amounted from the exact derivation of the low energy 
effective
action for the \N2 SYM theory \cite{SW1,SW2}. In principle, to get the exact
answer one needs to perform the summation of
infinite instanton series. Therefore, the result
by Seiberg and Witten was the first example of the exact derivation of the
total instanton sum. However, the authors did not calculate instantonic series,
using 
instead indirect and quite sophisticated methods. 
The calculation of the exact mass spectrum of
stable particles was their second essentially new result. 
What is important, this theory is non-trivial in one loop and
belongs to the class of asymptotically free theories.
In fact, the situation is a little bit unusual, since, despite the fact that
the direct instanton summation has not been performed yet, 
there are no doubts in the results derived which
survives under different tests. 

In order to get
the effective action unambiguously, the authors of \cite{SW1,SW2} used the 
three ideas: holomorphy \cite{shifmanvai}, 
duality \cite{montonen} and their compatibility with the renormalization group
flows.
Holomorphy implies that the low energy effective
action can be only of the form $\int{\Im \cal{F}}(\Psi)$ depending
on a single
holomorphic function $\cal{F}$ called prepotential. The duality
principle which provides the relation between the strong and week coupling 
regimes
is natural in the UV finite\footnote{In fact, one should require more: 
not renormalized and, therefore, non-running coupling constants.}, 
say ${\cal N}=4$ SUSY \cite{wittenolive}, theory, where the evident
modular parameter built from the coupling constant and the $\theta$-term

\be
\tau=\frac{4\pi i}{g^{2}}+{\theta\over 2\pi},
\ee
is not renormalized  and coincides with its asymptotic value. In
the theory with non-trivial renormalizations, the duality transformation 
does no longer interchange the weak and strong coupling regimes of the
same theory. Instead, it is much similar to the usual electric-magnetic
duality
and interchanges the low energy regimes of different theories. 
To construct this transformation explicitly, 
one needs an artificial trick, to introduce into the game
an additional object, a higher genus Riemann surface
whose period matrix coincides with the matrix of couplings
depending on the values of condensate in a given vacuum.

In the \N2 SYM theory this scheme is roughly 
realized as follows \cite{SW1,SW2}. 
First of all, in the Lagrangian there is a 
potential term for the scalar fields of the form

\be
V(\phi)=\Tr[\phi,\phi^{+}]^{2},
\ee
where the trace is taken over adjoint representation of the gauge group
(say, $SU(N_c)$). This potential gives rise to the valleys in the theory, when
$[\phi,\phi^+]=0$. 
The vacuum energy vanishes along the valleys, hence, 
the supersymmetry remains unbroken. One may always choose the v.e.v.'s of the 
scalar field to
lie in the Cartan subalgebra $\phi=diag(a_{1},...,a_{n})$. These parameters
$a_i$ can not serve, however, good order parameters, since there is still
a residual Weyl symmetry which changes $a_i$ but leave the same vacuum state. 
Hence, one should consider the set of the gauge invariant 
order parameters $u_{k}=<\Tr\phi^{k}>$ that fix the vacuum state unambiguously.
Thus, we obtain a moduli space parametrized
by the vacuum expectation values of the scalars, which is known
as the Coulomb branch of the whole moduli space of the theory. 
The choice of the point on the Coulomb branch is
equivalent to the  choice of the vacuum state and simultaneously
yields the scale which the coupling is frozen on. At the generic point of
the moduli space, the theory becomes effectively abelian after the
condensation of the scalar.

As soon as the scalar field acquires the vacuum expectation value, the
standard
Higgs mechanism works and there emerge heavy gauge bosons at large values
of the vacuum condensate. To derive the effective low energy action one has
to sum up the loop corrections as well as the multiinstanton
contributions. The loop corrections are trivial, since they are vanishing
beyond the one loop due to the supersymmetry. 
The explicit summation over instantons has not been done yet,
however, additional arguments allow one to define the effective action
indirectly.

The initial action of \N2 theory written in \N2 superfields has the simple
structure $S(\Psi )=\Im\tau \int \Tr\Psi^{2}$ with 
$\tau=\frac{4\pi i}{g^{2}} + {\theta\over 2\pi}$. Now the \N2 supersymmetry 
implies
that the low energy effective action gets renormalized only by holomorphic
contributions so that it is ultimately given by a single function
known as prepotential $S_{eff}(\Psi )=\Im\int{\cal{F}}(\Psi)$. The
prepotential
is a holomorphic function of moduli $u_k$ (or v.e.v.'s) 
except for possible singular points at the values of moduli where additional
massless states can appear disturbing the low energy behaviour.

Thus, the problem effectively reduces to the determination of one 
holomorphic function. If one manage to fix its behaviour nearby singularities,
the function can be unambiguously restored. One of the singularities, 
corresponding
to large values of v.e.v.'s, i.e. to the perturbative limit is under control
(since the theory is asymptotically free). All other singularities are treated
with the use of duality and of the non-renormalization theorems 
for the central charges of the SUSY algebra. A combination of these two ideas
allows one to predict the spectrum of the stable BPS states which become 
massless
in the deep non-perturbative region and are in charge of all other 
singularities.

The duality transformation can be easily defined
in the finite ${\cal N}=4$ SUSY theory just as the modular transformations 
generated by $\tau \to \tau^{-1}$ and $\tau \to \tau +1$. 
This makes a strong hint that 
the duality can be related with a modular space of some Riemann surfaces,
where
the modular group acts.
In fact, the naive application of duality meets some serious difficulties.
The reason is that, in the asymptotically free
theory, one has to match the duality with the renormalization 
group\footnote{In 
fact,
there are more problems with duality even in \N2 SUSY theories. In particular,
it unifies into one duality multiplet the monopole and gauge boson 
supermultiplets, 
while they have different spins, see, e.g., \cite{SW1,bilal}.}. This is 
non-trivial,
since now
$\tau$ depends on the scale which is supposedly involved into the duality 
transformation. The solution of this problem is that one is still able 
to connect the duality and 
modular transformations if considering {\it different low energy} theories
connected by the duality. Then, the duality acts on the moduli space
of vacua and this moduli space is associated with the moduli space of an
auxiliary Riemann surface, where the modular transformations act. 

At the next step, one has to find out proper variables whose modular
properties fit the field theory interpretation. These variables are 
the integrals\footnote{
We define the symbols $\oint$ and $res$ with
additional factors $(2\pi i)^{-1}$ so that
$$
{\res}_0 \frac{d\xi}{\xi}
= -{\res}_\infty \frac{d\xi}{\xi} =
\oint \frac{d\xi}{\xi} = 1
$$
} of a meromorphic 1-form $dS$
over the cycles on the Riemann surface, 
$a_i$ and $a^D_i$

\be\label{aad}
a_{i}=\oint_{A_{i}}dS, \\
a_{D}=\oint_{B_{i}}dS, 
\ee
(where $i,j=1,....,N_{c}-1$ for the gauge group $SU(N_{c})$).

These integrals play the two-fold role in the Seiberg-Witten approach.
First of all, one may calculate the prepotential ${\cal F}$ and, therefore, 
the low
energy effective action through the identification of
$a_D$ and $\partial \CF/\partial a$ with
$a$ defined as a function of moduli (values of condensate) by formula 
(\ref{aad}).
Then, using the property of the differential $dS$ 
that its variations w.r.t. moduli are holomorphic one may also calculate the
matrix of coupling constants

\be\label{Tij}
T_{ij}(u)=\frac{\partial^{2}{\cal{F}}}{\partial a_{i} \partial a_{j}},
\ee

The second role of formula (\ref{aad}) is that, as
was shown these integrals define the spectrum of the stable states
in the theory which saturate the Bogomolny-Prasad-Sommerfeld (BPS)
limit. For instance, the formula for the BPS spectrum in the $SU(2)$
theory reads as

\be
M_{n,m}=\left|na(u) +ma_{D}(u)\right|,
\ee
where the quantum numbers $n,m$ correspond to the ``electric" and ``magnetic"
states. The reader should not be confused with the spectrum derived
from the low energy behaviour which fixes arbitrarily heavy
BPS states. The point is that the BPS spectrum is related to the central
charge
of the extended SUSY algebra $Z_{\{m\},\{n\}}=a_in^i+a^D_im^i$
\cite{3a} and, therefore, has an anomaly origin.
On the other hand, anomalies are not renormalized by the quantum
corrections and can be evaluated in either of the UV and IR regions.

Note that the column $(a_i,a^D_i)$ transforms under
the action of the modular group $SL(2,\Z)$ as a section of the linear bundle.
Its global behaviour, in particular, the structure of the singularities 
is uniquely determined by the monodromy data. As we discussed earlier,
the duality transformation connects different singular points. In particular,
it interchanges ``electric", $a_i$ and ``magnetic", $a^D_i$
variables which describe the perturbative degrees of freedom at the
strong and weak coupling regimes respectively. 
Manifest calculations with the Riemann surface allow one to analyze
the monodromy properties of dual variables when
moving in the space of the order parameters. For instance,
in the simplest $SU(2)$ case, on the $u$-plane of the single order parameter 
there are three singular points, and the magnetic and
electric variables mix when encircling these points. Physically, 
in the theory with non-vanishing $\theta$-term (not pure imaginary $\tau$), 
the monopole acquires the electric
charge, while the polarization of the instanton medium yields the
induced dyons.

\section{\SW and theory of prepotential}

Although the both auxiliary objects, the Riemann surface and the differential 
$dS$ have come artificially, an attempt to recognize them
in SUSY gauge theories results into discovery of the integrable 
structures responsible for the Seiberg-Witten solutions.
 
Now let us briefly formulate the structures underlying Seiberg-Witten
theory. Later on, we often
refer to \SW as to the following set of data:
\begin{itemize}
\item
Riemann surface ${\cal C}$
\item
moduli space ${\cal M}$ (of the curves ${\cal C}$), the moduli space
of vacua of the gauge theory 
\item
meromorphic 1-form $dS$ on ${\cal C}$
\end{itemize}
How it was pointed out in \cite{GKMMM},
this input can be naturally described
in the framework of some underlying integrable system. 

To this end, first, we introduce bare spectral curve $E$ that is torus
$y^2=x^3+g_2x^2+g_3$ for the UV-finite
gauge theories with the associated holomorphic 1-form
$d\omega=dx/y$. This bare spectral curve degenerates into the
double-punctured sphere (annulus) for the asymptotically free theories
(where dimensional transmutation occurs): $x\to
w+1/w$, $y\to w-1/w$, $d\omega=dw/w$.
On this bare curve, there are given a
matrix-valued Lax operator $L(x,y)$. The corresponding dressed spectral
curve ${\cal C}$ is defined from the formula $\det(L-\lambda)=0$.

This spectral curve is a
ramified covering of $E$ given by the equation

\be
{\cal P}(\lambda;x,y)=0
\ee
In the case of the gauge group  $G=SU(N_c)$, the function ${\cal P}$ is a
polynomial of degree $N_c$ in $\lambda$.

Thus, we have the spectral curve ${\cal C}$, the moduli space ${\cal M}$ of the
spectral curve being given just  by
coefficients of ${\cal P}$.
The third important ingredient of the construction is the 
generating 1-form $dS \cong \lambda d\omega$ meromorphic on
${\cal C}$ (``$\cong$" denotes the equality modulo total derivatives). 
{}From the point of view of
the integrable system, it is just the shortened action "$pdq$" along the
non-contractible contours on the Hamiltonian tori. This means that the
variables $a_i$ in (\ref{aad}) are nothing but the action variables in the
integrable system. The defining property of $dS$ is
that its derivatives with respect to the moduli (ramification points)
are holomorphic differentials on the spectral curve.  This, in particular, 
means
that

\be
{\partial dS\over\partial a_i}=d\omega_i
\ee
where $d\omega_i$ are the canonical holomorphic differentials\footnote{I.e.
satisfying the conditions
$$
\oint_{A_i}d\omega_j=\delta_{ij},\ \ \ \ \oint_{B_i}d\omega_j=T_{ij}
$$
}.
Integrating this formula over $B$-cycles and using that
$a_D=\partial \CF/\partial a$, one immediately obtains (\ref{Tij}).

So far we reckoned without matter hypermultiplets.
In order to include them, one just needs to consider the
surface ${\cal C}$ with punctures. Then, the masses are proportional to the
residues of $dS$ at the punctures, and the moduli space has to be extended to
include these mass moduli. All other formulas remain in essence the same
(see \cite{WDVVlong} for more details).

The prepotential ${\cal F}$
and other ``physical" quantities are defined in terms of the
cohomology class of $dS$, formula (\ref{aad}). Note that
formula (\ref{Tij}) allows one to identify the prepotential with logarithm of 
the
$\tau$-function of the Whitham hierarchy \cite{typeB}: ${\cal F}=\log\tau$.

\section{\SW and integrable systems}

Thus, we have fixed
the set of data, that is, a Riemann surface,
the corresponding moduli space and the differential $dS$ that gives the
physical quantities, in particular, prepotential, and explained how it can be
associated with an integrable 
system. In the case of the $SU(2)$ pure gauge theory, which
has been worked out in detail in \cite{SW1} this correspondence can be 
explicitly checked \cite{GKMMM}. 
In particular, one may manifestly check formulas (\ref{aad}),
since all necessary ingredients are obtained in \cite{SW1}. In this way, one 
can
actually prove that this \SW corresponds to the integrable system (periodic 
Toda chain with two particles, see below). However, to repeat 
the whole Seiberg-Witten procedure for theories with more vacuum moduli
 is technically quite tedious if possible at all. Meanwhile, only in this way
one could unquestionably prove the correspondence in other cases. 

In real situation, the identification between the \SW and 
the integrable system
 comes via comparing the three characteristics
\begin{itemize}
\item number of the vacuum moduli and external parameters
\item perturbative prepotentials 
\item deformations of the two theories
\end{itemize}
The number of vacuum moduli (i.e. the number of scalar fields that may 
have non-zero v.e.v.`s) on the physical side should be compared with dimension 
of
the moduli space of the spectral curves in integrable systems, 
while the external parameters in
gauge theories (bare coupling constant, 
hypermultiplet masses) should be also some external parameters 
in integrable models (coupling constants, values of Casimir functions for
spin chains etc). 

As for second item, we already pointed out the distinguished role of
the prepotential in Seiberg-Witten theory (which celebrates a lot of 
essential properties, see e.g. \cite{mmm,RG}). 
In the prepotential, the contributions of particles and
solitons/mo\-no\-po\-les (dyons) sharing the same mass scale, are still
distinguishable, because of different dependencies on the bare coupling
constant, {\it i.e.} on the modulus $\tau$ of the bare coordinate elliptic
curve (in the UV-finite case) or on the $\Lambda_{QCD}$ parameter (emerging
after dimensional transmutation in UV-infinite cases).  In the limit $\tau
\rightarrow i\infty$ ($\Lambda_{QCD} \rightarrow 0$), the solitons/monopoles
do not contribute and the prepotential reduces to the ``perturbative'' one,
describing the spectrum of non-interacting {\it particles}.  It is
immediately given by the SUSY Coleman-Weinberg formula \cite{WDVVlong}:

\be\label{ppg}
{\cal F}_{pert}(a) \sim
\sum_{\hbox{\footnotesize reps}
\ R,i} (-)^F {\rm Tr}_R (a + M_i)^2\log (a + M_i)
\ee
Seiberg-Witten theory (actually, the identification of appropriate integrable
system) can be used to construct the non-perturbative prepotential,
describing the mass spectrum of all the ``light'' (non-stringy) excitations
(including solitons/monopoles).
Switching on Whitham times \cite{RG} presumably allows one to extract
some correlation functions in the ``light'' sector.

The perturbative prepotential (\ref{ppg}) has to be, certainly, added with
the classical part,

\be
\CF_{class}=-\tau a^2
\ee
since the bare action is $S(\Psi )=\Im\tau \int \Tr\Psi^{2}$. Here $\tau$ is
the value of the coupling constant in the deeply UV region. It plays
the role of modular parameter of the bare torus in the UV finite theories. 
On the other hand, in asymptotically free theories, where the perturbative
correction behaves like $a^2\log (a/\Lambda_{QCD})$, $\tau$ is just contained
in $\Lambda_{QCD}$, since it can be invariantly defined neither in integrable,
nor
in gauge theories.

In fact, the problem of calculation of the prepotential in physical
theory is simple only at the perturbative level, where it is given just by
the leading contribution,
since the $\beta$-function in ${\cal N}=2$ theories is non-trivial only in one
loop.
However, the calculation of all higher (instantonic) corrections in the gauge 
theory can be hardly
done at the moment\footnote{In order to check the very ideology that
the integrable systems lead to the correct answers, there were
calculated first several corrections \cite{Instanton}. 
The results proved to exactly 
coincide with the predictions obtained within the integrable approach.}.
Therefore, the standard way of doing is to make an identification
of the \SW and an integrable system and then to rely on integrable 
calculations.
This is why establishing the ``gauge theories $\leftrightarrow$
integrable theories" correspondence is of clear practical 
(apart from theoretical) importance. 

Now let us come to the third item in the above correspondence, namely, how one
can
extend the original Seiberg-Witten theory. There are basically three
different ways to extend the original \SW.

First of all, one may consider other gauge groups, from other simple 
classical groups to those being a product of several simple factors.
The other possibility is to add some matter hypermultiplets in different
representations. The two main cases here are the matter in fundamental or
adjoint representations. At last, the third possible direction
to deform \SW is to consider 5- or 6-dimensional theories, compactified
respectively onto the circle of radius $R_5$ or torus with modulus $R_5/R_6$ 
(if the number of dimensions exceeds 6, the gravity becomes 
obligatory coupled to the gauge theory).

Now we present a table of 
known relations between gauge theories and integrable systems, see Fig.1.

\begin{figure}
%[t]
\begin{center} \begin{tabular}{|c|c|c|c|c|}
\hline
{\bf SUSY}&{\bf Pure gauge}&{\bf SYM theory}&{\bf SYM theory}\\
{\bf physical}&{\bf SYM theory,}&{\bf with adj.}&{\bf with fund.}\\
{\bf theory}&{\bf gauge group $G$}&{\bf matter}&{\bf matter}\\
\hline
        & inhomogeneous  & elliptic  & inhomogeneous        \\
         & periodic        & Calogero & periodic \\
{\bf 4d} & Toda chain & model & \XXX \\
& for the dual affine ${\hat G}^{\vee}$  &({\it trigonometric}& spin chain\\
        & ({\it non-periodic}  & {\it Calogero}& ({\it non-periodic}\\
        & {\it Toda chain})& {\it model})& {\it chain})\\
\hline
        & periodic  & elliptic & periodic   \\
        & relativistic  & Ruijsenaars&\XXZ\\
{\bf 5d} & Toda chain  & model & spin chain  \\
        & ({\it non-periodic} & ({\it trigonometric}& ({\it non-periodic}\\
        & {\it chain}) & {\it Ruijsenaars})&{\it chain})\\
\hline
    &periodic & Dell & periodic\\
    & ``Elliptic" & system & \XYZ (elliptic)\\
{\bf 6d} &  Toda chain & ({\it dual to elliptic} & spin chain\\
        & ({\it non-periodic}& {\it Ruijsenaars,}&({\it non-periodic} \\
& {\it chain})& {\it elliptic-trig.})&{\it chain})\\
\hline
\end{tabular}
\end{center}
\caption{SUSY gauge theories $\Longleftrightarrow$ integrable systems
correspondence. The perturbative limit is marked by the italic font (in
parenthesis).}\label{intvsYM}
\end{figure}
\vspace{10pt}

Let us briefly describe different cells of the table.

The original Seiberg-Witten model, which is the $4d$ pure gauge $SU(N)$ theory
(in fact, in their papers \cite{SW1,SW2}, the authors considered the $SU(2)$
case only, but the generalization made in \cite{3a} is quite
immediate), is the upper left square of the table. The
remaining part of the table contains possible deformations.
Here only two of the three possible ways to deform the original Seiberg-Witten
model are shown. Otherwise, the table would be three-dimensional. 
In fact, the third direction related to changing the gauge group,
although being of an interest is slightly out of the main line.  

One direction in the table corresponds to matter hypermultiplets
added. The most interesting is to add matter in adjoint or fundamental 
representations, although attempts to add antisymmetric and
symmetric matter hypermultiplets were also done (see \cite{asym} for
the construction of the curves and \cite{KPasym} for the corresponding
integrable systems). Adding matter in other representations in the basic
$SU(N)$ case leads to non-asymptotically free theories.

\paragraph{Columns: Matter in adjoint {\it vs.} fundamental 
representations of the gauge group.}  

Matter in adjoint representation can be described in terms of a
larger pure SYM model, either with higher SUSY or in higher dimensional 
space-time.
 Thus models with adjoint matter form a hierarchy, naturally
associated with the hierarchy of integrable models {\it Toda chain
$\hookrightarrow$ Calogero $\hookrightarrow$
Ruijsenaars $\hookrightarrow$ Dell}
\cite{GKMMM,dw,IM,bmmm2,bmmm3,MM1,MM2}. Similarly, the
models with fundamental matter are related to the hierarchy of spin chains
originated from the Toda chain: {\it Toda chain
$\hookrightarrow$ \XXX $\hookrightarrow$
\XXZ $\hookrightarrow$ \XYZ}
\cite{XXX}.

Note that, while coordinates in integrable systems describing pure gauge
theories and those with fundamental matter, live on the cylinder (i.e.
the dependence on coordinates is trigonometric), the coordinates in the
Calogero system (adjoint matter added)
live on a torus\footnote{Since these theories are
UV finite, they depend on an additional (UV-regularizing) parameter, which
is exactly the modulus $\tau$ of the torus.}. However, when one takes the
perturbative limit, the coordinate dependence becomes trigonometric.

\paragraph{Lines: Gauge theories in different dimensions.}

Integrable systems relevant
for the description of vacua of $d=4$ and $d=5$ models are
respectively the Calogero and Ruijsenaars ones (which possess the ordinary
Toda chain and ``relativistic Toda chain'' as Inozemtsev's limits
\cite{Ino}), while $d=6$ theories are described by the
double elliptic (Dell) systems. When we go from $4d$ (Toda, \XXX, Calogero) 
theories to (compactified onto circle) 
$5d$ (relativistic Toda, \XXZ, Ruijsenaars) theories 
the momentum-dependence of the Hamiltonians
becomes trigonometric (the momenta live on the compactification circle)
instead of rational. Similarly, the (compactified onto torus) $6d$ theories 
give rise to an elliptic momentum-dependence of the Hamiltonians, with
momenta living on the compactification torus. Since adding
the adjoint hypermultiplet elliptizes the coordinate dependence, the
integrable
system corresponding to $6d$ theory with adjoint\footnote{Let us point out 
that by adding the adjoint matter we always mean soft breaking of 
higher supersymmetries.} 
matter celebrates both
the coordinate- and momentum-dependencies elliptic. A candidate for ``the
elliptic Toda chain'' was proposed in \cite{Kriel}.

Further in this part we show how the Riemann surfaces and the meromorphic
differentials arise when considering a concrete integrable many-body system.
We are going to explain in more details the simplest top left corner of the 
table
and make some comments on the adjoint matter case in different dimensions.
Then, after discussing duality, we shall come to the most general adjoint 
matter system,
the right bottom cell in the table which is our ultimate target in these 
lectures.
It can be, however, done only with the use of the coordinate-momentum 
duality\footnote{This duality has nothing to do with the duality of 
\cite{SW1,SW2} 
which we discussed earlier.}.

\section{$4d$ pure gauge theory: Toda chain}

Thus, we start with the simplest case of the $4d$ pure gauge theory, which was
studied in details in the original paper \cite{SW1} for the $SU(2)$ gauge
group and straightforwardly generalized to the $SU(N_c)$ case in \cite{3a}.

This system is described by the periodical Toda chain
with period $N_c$ \cite{GKMMM,MW,6} whose equations of motion read

\be\label{Todaeq}
\frac{\partial q_i}{\partial t} = p_i \ \ \ \ \
\frac{\partial p_i}{\partial t} = e^{q_{i+1} -q_i}-
e^{q_i-q_{i-1}}
\ee
with periodic boundary conditions $p_{i+N_c}=p_i$, $q_{i+N_c}=q_i$ imposed.
In physical system, there are $N_c$ moduli and there are no external
parameters ($\Lambda_{QCD}$ can be easily removed by rescaling).
In the Toda system, there are exactly $N_c$ conservation laws.
These conservation laws can be constructed from the Lax operator 
defined for any integrable system. The eigenvalues of the Lax operator
do not evolve, thus, any function of the eigenvalues is an integral of motion.

There are two different Lax representation describing the periodic Toda chain.
In the first one, the Lax operator is represented by the $N_c\times N_c$
matrix depending on dynamical variables 

\be\label{LaxTC}
{\cal L}^{TC}(w) =
\left(\begin{array}{ccccc}
p_1 & e^{{1\over 2}(q_2-q_1)} & 0 & & we^{{1\over 2}(q_1-q_{N_c})}\\
e^{{1\over 2}(q_2-q_1)} & p_2 & e^{{1\over 2}(q_3 - q_2)} & \ldots & 0\\
0 & e^{{1\over 2}(q_3-q_2)} & -p_3 & & 0 \\
 & & \ldots & & \\
\frac{1}{w}e^{{1\over 2}(q_1-q_{N_c})} & 0 & 0 & & p_{N_c}
\end{array} \right)
\ee
The characteristic equation for the Lax matrix

\be\label{SpeC}
{\cal P}(\lambda,w) = \det_{N_c\times
N_c}\left({\cal L}^{TC}(w) - \lambda\right) = 0
\ee
generates the conservation laws and 
determines the spectral curve

\be\label{fsc-Toda}
w + \frac{1}{w} = 2P_{N_{c}}(\lambda ) 
\ee
where $P_{N_c}(\lambda)$ is a polynomial of degree $N_c$ whose
coefficients are integrals of motion. 

The spectral curve (\ref{fsc-Toda}) is exactly the Riemann surface 
introduced in the context of $SU(N_c)$ gauge theory. 
The integrals of motion parametrize the moduli space of the
complex structures of the hyperelliptic surfaces of genus $N_c-1$, which is
the moduli space of vacua in physical theory. 

If one considers the case of two particles ($SU(2)$ gauge theory), $P(\lambda)
=\lambda^2-u$, $u=p^2-\cosh q$, $p=p_1=-p_2$, $q=q_1-q_2$.
Thus we see that the order parameter
of the SUSY theory plays the role of Hamiltonian 
in integrable system. In the perturbative regime of the
gauge theory, one of the exponentials in $\cosh q$ vanishes, and one
obtains the non-periodic Toda chain. In the equation (\ref{fsc-Toda}) the 
perturbative limit implies vanishing the second term in the l.h.s., i.e.
the spectral curve becomes the sphere $w=2P_{N_c}(\lambda)$.

After having constructed the Riemann surface and the moduli space 
describing the $4d$ pure gauge theory, 
we turn to the third crucial ingredient of \SW
that comes from integrable systems, the generating differential $dS$.
The general construction was explained in the Introduction, 
this differential is
in essence the ``shorten" action $pdq$. Indeed, in order to construct 
action variables, $a_i$ 
one needs to integrate the differential ${\tilde {dS}}=\sum_ip_idq_i$
over $N_c-1$ non-contractable cycles in the Liouville torus which is nothing but
the level submanifold of the phase space, i.e. the submanifold defined by
values of all $N_c-1$ integrals of motion fixed. On this submanifold, the
momenta $p_i$ are functions of the
coordinates, $q_i$. The Liouville torus in \SW
is just the Jacobian
corresponding to the spectral curve (\ref{fsc-Toda}) 
(or its factor over a subgroup) .

In the general case of a $g$-parameter family of complex curves
(Riemann surfaces) of genus $g$, the Seiberg-Witten differential
$dS$ is characterised by the property
$\delta dS = \sum_{i=1}^g \delta u_i dv_i$,
where $dv_i(z)$ are $g$ holomorphic 1-differentials
on the curves (on the fibers), while $\delta u_i$ are
variations of $g$ moduli (along the base).
In the associated integrable system, $u_i$ are integrals of motion
and $\pi_i$, some $g$ points on the curve are momenta.
The symplectic structure is

\be\label{dSjac}
\sum_{i=1}^g da_i\wedge dp_i^{Jac} = \sum_{i,k=1}^g
du_i\wedge dv_i(\pi_k)
\ee
The vector of the angle variables,

\be\label{pjac}
p_i^{Jac} = \sum_{k=1}^g \int^{\pi_k} d\omega_i
\ee
is a point of the Jacobian, and the Jacobi map identifies this with the
$g$-th power of the curve, $Jac\ \cong {\cal C}^{\otimes g}$.
Here $d\omega_i$ are {\it canonical} holomorphic differentials,
$dv_i = \sum_{j=1}^g d\omega_j \oint_{A_j} dv_i$.
Some details on the
symplectic form on the finite-gap solutions can be found in \cite{konzon}.

Technical calculation is, however, quite tedious. It is simple only in the
2-particle ($SU(2)$) case, when the Jacobian coincides with the curve itself.
In this case, the spectral curve is 

\be\label{sc2}
w+{1\over w}=2(\lambda^2-u)
\ee
while $u=p^2-\cosh q$. Therefore, one can write for the action variable

\be\label{dSTC}
a=\oint pdq=\oint \sqrt{u-\cosh q}dq=\oint \lambda {dw\over w},\ \ \ 
dS=\lambda {dw\over w}
\ee
where we made the change of variable $w=e^q$ and used equation (\ref{sc2}).

Now one can naturally assume that this expression for the differential $dS$
is suitable for generic $N_c$. A long calculation (which can be borrowed, say,
from the book by M.Toda \cite{toda}) shows that this is really the 
case. One can easily check that the derivatives of $dS$ w.r.t. to moduli are 
holomorphic, up to total derivatives. Say, if one 
parametrizes\footnote{The absence of the term $\lambda^{N_c-1}$ is due
to the $SU(N_c)$ group and corresponds to the total momentum equal to zero,
i.e. to the centre mass frame. This is important for the counting of 
genus.} 
$P_{N_c}(\lambda)=-\lambda^{N_c}+s_{N_c-2}\lambda^{N_c-2}+...$ and note
that $dS=\lambda dw/w=\lambda dP_{N_c}(\lambda)/Y=P_{N_c}(\lambda) d\lambda/Y
+ \hbox{ total derivatives }$, then

\be
{\partial dS\over\partial s_k}={\lambda^kd\lambda\over Y}
\ee 
and these differentials are holomorphic if $k\le N_c-2$. 
Thus, $N_c-1$ moduli gives rise to $N_c-1$
holomorphic differentials which perfectly fits the genus of the curve (we
use here that there is no modulus $s_{N_c-1}$).

\section{Perturbative prepotentials}
\setcounter{equation}{0}

In the previous section, we established in the concrete example
the connection between the Seiberg-Witten 
integrable theory and the gauge theory. However, so far we studied only 
a kind of
general landscape in the theory, since the real, numerical results enter the
game only with the prepotential calculated. Moreover, the correspondence
between gauge and integrable theories is supported by comparing the 
perturbative
prepotentials. We discuss it now.

The technical tool that allows one to proceed with effective 
perturbative expansion of the prepotential is ``the residue formula'',
the variation of the period matrix, i.e. the third derivatives
of ${\cal F}(a)$ (see, e.g. \cite{WDVVlong}):

\be
\frac{\partial^3{\cal F}}{\partial a_i\partial a_j\partial a_k}
= \frac{\partial T_{ij}}{\partial a_k} =
{\rm res}_{d\xi = 0} \frac{d\omega_id\omega_jd\omega_k}{\delta dS},
\label{res}
\ee
where $\delta dS\equiv d\left({dS\over d\xi}\right)d\xi $, or, explicitly
$\delta dS=d\lambda d\xi$. 
Here the differential\footnote{
We reserve notation $\xi$ for the variable living on the bare spectral curve 
(torus), while shall denote later through 
$\zeta$ the variable living on the compactification torus
in 6 dimensions.} 
$d\xi$ given on the bare torus 
degenerates into $dw/w$ when torus degenerates into the punctured sphere
(Toda chain).
We remark that although
$d\xi$ does not have zeroes on the {\it bare} spectral
curve when it is a torus or doubly punctured sphere, it does
in general however possess them on the covering ${\cal C}$. 

In order to construct the perturbative prepotentials, one merely 
can note that, at the leading order, the Riemann surface (spectral
curve of the integrable system) becomes rational. Therefore, the
residue formula allows one to obtain immediately the third derivatives
of the prepotential as simple residues on sphere. In this way, 
one can check
that the prepotential has actually the form (\ref{ppg}) \cite{WDVVlong}.
Later we discuss these calculations, but right now 
we shall see what the perturbative prepotentials are in 
in gauge theory.

As a concrete example, let us consider the $SU(N_c)$ gauge group. Then, say,
the perturbative prepotential for the pure gauge theory acquires the
form

\be\label{pppg}
{\cal F}_{pert}^V={\f 8\pi i}\sum_{ij}
\left(a_i-a_j\right)^2\log\left(a_i-a_j\right)
\ee
This formula establishes that when v.e.v.'s
of the scalar fields in the gauge supermultiplet are non-vanishing
(perturbatively $a_r$ are eigenvalues of the vacuum
expectation matrix  $\langle\phi\rangle$), the fields in the gauge multiplet
acquire masses $m_{rr'} = a_r - a_{r'}$ (the pair of indices $(r,r')$ label
a field in the adjoint representation of $SU(N_c)$). 
The eigenvalues are subject to the condition $\sum_ia_i=0$.
Analogous formula for the
adjoint matter contribution to the prepotential is

\be\label{pppa}
{\cal F}_{pert}^A=-{\f 8\pi i}\sum_{ij}
\left(a_i-a_j+M\right)^2\log\left(a_i-a_j+M\right)
\ee

These formulas can be almost immediately extended to the
``relativistic"
 $5d$ \N2 SUSY gauge models
with one compactified dimension. One can understand the
reason for this ``relativization'' in the following way. Considering
four plus one compact dimensional theory one should take into account the
contribution of all Kaluza-Klein modes to each 4-dimensional field.
Roughly speaking it leads to the 1-loop contributions to the effective
charge of the form ($a_{ij}\equiv a_i-a_j$)

\be\label{relKK}
T_{ij} \sim \sum _{\rm masses}\log\hbox{ masses} \sim \sum _m\log
\left(a_{ij} +
{m\over R_5}\right) \sim \log\prod _m\left(R_5a_{ij} + m\right)
\sim\log\sinh R_5a_{ij}
\ee
i.e. coming from $4d$ to $5d$
one should make a substitution $a_{ij} \rightarrow 
\sinh R_5a_{ij}$, at least, in the
formulas for perturbative prepotentials. Adding the adjoint matter in this 
case would
give the effective charge (or, equivalently, the period matrix of the spectral
curve
of the \RS system \cite{bmmm2}, see table 1): 

\be\label{3.67}
T_{ij}\sim\log\left({\sqrt{\sinh\left(R_5a_{ij}+\epsilon\right)
\sinh\left(R_5a_{ij}-\epsilon\right)}
\over\sinh R_5a_{ij}}\right)
\ee
The same general argument can be equally applied to the $6d$ case,
or to the theory with {\em two} extra compactified dimensions,
of radii $R_5$ and $R_6$. Indeed, the account of
the Kaluza-Klein modes allows one to predict the perturbative form
of charges in the $6d$ case as well. Namely, one should expect them
to have the form\footnote{We denote through $\theta_*$ the 
$\theta$-function with
odd characteristics \cite{BEWW}.}

\be\label{6dKK}
T_{ij} \sim \sum _{\rm masses}\log\hbox{ masses} \sim \sum _{m,n}\log
\left(a_{ij} +
{m\over R_5} +{n\over R_6}
\right) \sim 
\\
\sim\log\prod _{m,n}\left(R_5a_{ij} + m+n{R_5\over R_6}\right)
\sim\log\theta_*\left(R_5a_{ij}\left|i{R_5\over R_6}\right)\right.
\ee
i.e. coming from $4d$ ($5d$) to $6d$
one should replace the rational (trigonometric) expressions by the elliptic 
ones, at least, in the
formulas for perturbative prepotentials. In this case adding the adjoint 
matter
would lead to the effective charge

\be\label{3.68}
T_{ij}\sim\log\left({\sqrt{\theta_*\left(R_5a_{ij}+\epsilon
\left|i{R_5\over R_6}\right)\right.
\theta_*\left(R_5a_{ij}-\epsilon\left|i{R_5\over R_6}\right)\right.}
\over \theta_*\left(R_5a_{ij}\left|i{R_5\over R_6}\right)\right.}\right)
\ee

\section{$4d$ theory: a simple example of calculation}

Here we describe the method of calculating prepotentials in the simplest
example of the Toda chain. In the perturbative limit 
(${\Lambda_{QCD}\over\langle\bphi\rangle}\rightarrow 0$) 
the second term in (\ref{fsc-Toda}) vanishes, the curve acquires the form 

\be\label{pertcurv} 
W = 2P_{N_c}(\lambda),\ \ \ P_{N_c}(\lambda)=\prod (\lambda -\lambda_i) 
\ee 
(a rational curve with punctures
\footnote{These punctures emerge as a 
degeneration of the handles of the hyperelliptic surface so that the 
$A$-cycles encircle the punctures.}) and the generating differential 
(\ref{dSTC}) turns into

\be\label{pertdS} 
dS^{(4)}_{\rm pert} = \lambda d\log P_{N_c}(\lambda)
\ee 

Now the set of the $N_c-1$ independent 
canonical holomorphic differentials is

\be\label{difff} 
d\omega_i=\left({1\over\lambda-\lambda 
_i}-{1\over\lambda -\lambda_{N_c}}\right)d\lambda= 
{\lambda_{iN_c}d\lambda\over 
(\lambda -\lambda_i)(\lambda -\lambda_{N_c})},\ \ \ i=1,...,N_c-1, 
\ \ \ \lambda_{ij}\equiv \lambda_i-\lambda_j 
\ee 
and the $A$-periods $a_i = \lambda_i$ (the independent ones are, say, with 
$i=1,\dots,N_c-1$) coincide with the roots of polynomial $P_{N_c}(\lambda)$, 
$\lambda_i$.
The prepotential can be now computed via the residue formula

\be 
\CF_{pert,ijk}=\stackreb{d\log {P_{N_c}}=0}{\res} 
{d\omega_id\omega_jd\omega_k\over d\log 
{P_{N_c}(\lambda)}d\lambda} 
\ee 
which can be explicitly done. The only technical trick is that it
is easier to compute the residues at the poles of $d\omega$'s instead of the
zeroes of $d\log P$. This can be done immediately since there are no 
contributions from the infinity $\lambda=\infty$. 
The final results have the following form 

\be  
2\pi i\CF_{pert,iii}=\sum_{k\ne i}{1\over \lambda_{ik}}+{6\over 
\lambda_{iN_c}}+ 
\sum_{k\ne N_c}{1\over \lambda_{kN_c}}, 
\\ 
2\pi i\CF_{pert,iij}= {3\over \lambda_{iN_c}}+{2\over 
\lambda_{jN_c}}+\sum_{k\ne i,j,N_c} 
{1\over \lambda_{kN_c}}-{\lambda_{jN_c}\over \lambda_{iN_c}\lambda_{ij}}, 
\ \ \ i\ne j,\\ 
2\pi i\CF_{pert,ijk}=2\sum_{l\ne N_c}{1\over \lambda_{lN_c}}-
\sum_{l\ne i,j,k,N_c} 
{1\over \lambda_{lN_c}}, 
\ \ \ i\ne j\ne k; 
\ee
giving rise to the prepotential 
formula 

\be\label{FA}
{\cal F}_{pert} = {1\over 8\pi i}\sum _{ij}f^{(4)}(\lambda_{ij}),
\ \ \ \
f^{(4)}(x) = x^2\log x^2
\ee 
This formula perfectly matches the perturbative prepotential expected 
in the gauge theory (\ref{pppg}).

\part{Dualities in the many-body systems and gauge theories}

In this part, we are going to discuss a property of some integrable systems
which is called duality. This property will allow us to fill in the 
right bottom
corner of table 1, i.e. to construct a double elliptic system.

In fact, as in string theory there are several dualities (S-, T- and 
U-dualities),
as in integrable systems there are several kinds of duality \cite{gnr}. 
Here we concentrate on the most interesting, important and still 
remaining mysterious one, the coordinate-momentum duality. This is a duality
between pairs of dynamical systems (which is self-duality for
some systems) \cite{rud,ruijh,ruo,fgnr,bmmm3}.
Note that this duality at the moment is attributed only with 
the Hitchin type systems, i.e. the Seiberg-Witten theories with adjoint 
matter, while 
spin chains 
are still not embedded into this treatment. 

Let us consider a system with the phase space
being a hyperkahler four dimensional manifold. Let this manifold
involve at most two tori, i.e. be the
elliptically fibered K3 manifold.
One torus is that where the momenta live and
the other one is that where the coordinates live. The duality at hands
actually interchanges momentum and coordinate tori and, therefore,
can be called $pq$-duality. Note that typically one deals with 
degenerations of these tori.

One can also consider a theory which is self-dual. This integrable system
is called Dell (double elliptic)
system\footnote{In fact, this is an example of the Dell system with 2 degrees
of freedom. One may equally consider the Dell system with arbitrary many
degrees of freedom, see below.} and
describes the Seiberg-Witten theory for the six-dimensional
theory with adjoint matter.

The other cases correspond to some degenerations of Dell system. 
Say, the degeneration
of the momentum torus to $C/Z_{2}$ leads
to the five-dimensional theory, while the degeneration to $R^2$
leads to the four-dimensional theory. Since the modulus
of the coordinate torus has meaning of the
complexified bare coupling constant in field theory, interpretation
of the degenerations of the coordinate torus is different.
In particular, the degeneration to the cylinder corresponds to switching
off the instanton effects, i.e. to the perturbative regime.

A generic description of the $pq$-duality 
expresses a relationship between two completely
integrable systems $S_1, S_2$ on a fixed symplectic manifold with given
symplectic structure $(M,\omega)$ and goes back to
\cite{rud}.
We say the Hamiltonian systems are {\it dual} when the conserved quantities of
$S_1$ and $S_2$ together form a coordinate system for $M$. Consider for
example free particles, $H^{(1)}_k=\sum_i p^k_i/k$. For this system the free
particles momenta are identical to the conserved quantities or action
variables. Now consider the Hamiltonian $H^{(2)}_k=\sum_i q^k_i/k$ with
conserved quantities $q_i$. Together $\{p_i,q_i\}$ form a coordinate system
for phase space, and so the two sets of Hamiltonians are dual. Duality then
in this simplest example is a transformation which interchanges momenta and
coordinates. For more complicated interacting integrable systems finding dual
Hamiltonians is a non-trivial exercise. Note that this whole construction
manifestly depends on the particular choice of conserved quantities. A
clever choice may result in the dual system arising by simply interchanging
the momentum and coordinate dependence, as in the free system.

Some years ago Ruijsenaars \cite{rud} observed such
dualities between various members of the Calogero-Moser and Ruijsenaars
families: the rational Calogero and trigonometric Ruijsenaars models were
dual to themselves while trigonometric Calogero model was dual with the
rational Ruijsenaars system (see \cite{ruo,fgnr} for more examples).
These dualities were shown by starting with a
Lax pair $L=L(p,q)$ and an auxiliary diagonal matrix $A=A(q)$. When $L$ was
diagonalized the matrix $A$ became the Lax matrix for the dual Hamiltonian,
while $L$ was a function of the coordinates of the dual system. Dual systems 
for
a model possessing a Lax representation are then related to the eigenvalue
motion of the Lax matrix. We discuss this approach in more details below.

Another approach to finding a dual system is to make a canonical
transformation which
substitutes the original set of Poisson-commuting coordinates $q_i$,
$\{q_i,q_j\} = 0$, by another obvious
set of the Poisson-commuting
variables: the Hamiltonians $h_i(\vec p,\vec q)$ or, better, the
action variables $a_i(\vec h) = a_i(\vec p,\vec q)$.
It will be clear below that in practice really interesting
transformations are a little more sophisticated: $h_i$ are identified with
certain functions of the new coordinates (these functions determine the
Ruijsenaars matrix $A(q)$), which -- in the most interesting cases --
are just the same Hamiltonians with the interactions switched-off. Such free
Hamiltonians are functions of momenta alone, and the dual coordinates 
substitute
these momenta, just as one had for the system of free particles.

The most interesting question for our purposes is: {\it what are the duals of
the elliptic Calogero and Ruijsenaars systems ?}
Since the elliptic Calogero (Ruijsenaars) is rational
(trigonometric) in momenta and elliptic in the coordinates, the dual
will be elliptic in momenta
and rational (trigonometric) in coordinates.
Having found such a model the final elliptization of the coordinate
dependence is straightforward, providing us with the wanted
double-elliptic systems.

\section{$pq$-duality: two-body system}

The way of constructing dual systems via a canonical transformation
is technically quite tedious for many-particle systems, but
perfectly fits the case of $SU(2)$
which, in the center-of-mass frame, has only one coordinate and
one momentum.
In this case the duality transformation can be described explicitly since
the equations of motion can be integrated in a straightforward way.
Technically, given two Hamiltonian systems, one with the momentum $p$,
coordinate $q$ and
Hamiltonian $h(p,q)$ and another with the momentum $P$,
coordinate $Q$ and
Hamiltonian $H(P,Q)$ we may describe duality by the relation

\be
h(p,q) = f(Q), \nn \\
H(P,Q) = F(q).
\ee
Here the functions $f(Q)$ and $F(q)$ are such that

\be
dP\wedge dQ = -dp\wedge dq,
\ee
which expresses the fact we have an (anti-)canonical transformation. This
relation entails that

\be
F'(q)\frac{\partial h(p,q)}{\partial p} =
f'(Q)\frac{\partial H(P,Q)}{\partial P}.
\label{pbconseq}
\ee

At this stage the functions $f(Q)$ and $F(q)$ are arbitrary.
However, when the Hamiltonians depend on a coupling constant $\nu^
2$
and are such that their ``free'' part can be separated and depends
only on the momenta,\footnote{
Note that this kind of duality relates the weak coupling
regime for $h(p,q)$ to the weak coupling regime for
$H(P,Q)$. For example, in the rational Calogero case
\be\label{ratCdual}
h(p,q) = \frac{p^2}{2} + \frac{\nu^2}{q^2} = \frac{Q^2}{2}\\
H(P,Q) = \frac{P^2}{2} + \frac{\nu^2}{Q^2} = \frac{q^2}{2}
\ee
We recall that, on the field theory side, the coupling constant
$\nu$ is related to the mass of adjoint hypermultiplet and
thus remains unchanged under duality transformations.
}
the free Hamiltonians provide a natural choice for these functions:
$F(q) = h_0(q)$ and $f(Q) = H_0(Q)$ where

\be
h(p,q)\big|_{\nu^2 = 0} = h_0(p), \\
H(P,Q)\big|_{\nu^2 = 0} = H_0(P).
\ee
With such choice the duality equations become

\be
h_0(Q) = h(p,q) \label{duality1},\\
H_0(q) = H(P,Q) \label{duality2},\\
\frac{\partial h(p,q)}{\partial p}H'_0(q) =
h'_0(Q)\frac{\partial H(P,Q)}{\partial P}.
\label{duality3}
\ee

Free rational,
trigonometric and elliptic Hamiltonians are
$h_0(p) = \frac{p^2}{2}$, $h_0(p) = \cosh p$ and $h_0(p) =\cn(p|k)$
respectively.

Note that from the main duality relation,

\be
H_0(q) = H(P,Q)
\label{duality}
\ee
it follows that 

\be
\left.\frac{\partial q}{\partial P}\right|_Q = \frac{1}{H_0'(q)}
\frac{\partial H(P,Q)}{\partial P}
\ee
which together with (\ref{duality3}) implies:

\be
\left.\frac{\partial q}{\partial P}\right|_Q =
\frac{1}{h_0'(Q)}\frac{\partial h(p,
q)}{\partial p}
\ee
When compared with the Hamiltonian equation for the
original system,

\be
\frac{\partial q}{\partial t} = \frac{\partial h(p,q)}{\partial p},
\ee
we see that $P = h'_0(Q)t$ is proportional to the ordinary time-variable $t$, 
while $h_0(Q)=h(p,q)=E$ expresses $Q$ as a function of the energy $E$.
This is a usual feature of classical integrable systems, exploited in
Seiberg-Witten theory \cite{GKMMM}:
in the $SU(2)$ case the spectral curve $q(t)$ can be described by

\be
h\left(p\left(\frac{\partial q}{\partial t},q\right),\ q\right) = E.
\label{times}
\ee
where $p$ is expressed through $\partial q/\partial t$ and $q$
from the Hamiltonian equation
$\partial q/\partial t = \partial H/\partial p$.
In other words, the spectral curve is
essentially the solution of the equation of motion of integrable system,
where the time $t$ plays the role of the spectral parameter
and the energy $E$ that of the modulus.

Let us consider several simple examples. The simplest one is
the rational Calogero. In this case, the duality transformation connects
two identical Hamiltonians, (\ref{ratCdual}). Therefore, this system is 
self-dual.
Somewhat less trivial example is the trigonometric Calogero-Sutherland 
model \cite{C,S}. It leads
to the following dual pair 

\be
h(p,q) = \frac{p^2}{2} + \frac{\nu^2}{\sin^2 q} = \frac{Q^2}{2}\\
H(P,Q) = \sqrt{1-{2\nu^2\over Q^2}}\cos P = \cos{q}
\ee
One may easily recognize in the second Hamiltonian the rational Ruijsenaars
Hamiltonian \cite{rru2,ruijrev}. Thus, the Hamiltonian rational in coordinates and 
momenta
maps onto itself under the duality, while that with rational momentum
dependence and trigonometric coordinate dependence maps onto the Hamiltonian
with inverse, rational coordinate and trigonometric momentum dependencies. 
Therefore, we can really see that the $pq$-duality exchanges
types of the coordinate and momentum dependence. This implies that the 
Hamiltonian
trigonometrically dependent both on coordinates and momenta is to be
self-dual. Indeed, one can check that the trigonometric
\RS system is self-dual:

\be
h(p,q) = \sqrt{1-{2\nu^2\over q^2}}\cos p = \cos{Q}\\
H(P,Q) = \sqrt{1-{2\nu^2\over Q^2}}\cos P = \cos{q}
\ee
Now we make the next step and introduce into the game elliptic 
dependencies.

\section{Two-body system dual to Calogero and Ruijsenaars}

We begin with the elliptic Calogero Hamiltonian\footnote{The standard
definitions of the Jacobi functions we use below can be found in 
\cite{BEWW}.} \cite{C,M}

\be
h(p,q) = \frac{p^2}{2} + \frac{\nu^2}{\sn^2(q|k)},
\ee
and seek a dual Hamiltonian elliptic in the momentum. Thus
$h_0(p)=\frac{p^2}{2}$ and we seek $H(P,Q)=H_0(q)
$ such that $H_0(q) =
\cn(q|k)$. Eqs.(\ref{duality}) become

\be
\frac{Q^2}{2} = \frac{p^2}{2} + \frac{\nu^2}{\sn^2(q|k)},\ \ \ 
\cn(q|k) = H(P,Q), \ \ \ 
p \cdot \cn'(q|k) = Q\frac{\partial H(P,Q)}{\partial P}.
\label{Calduality}
\ee
Upon substituting

\be
\cn'(q|k) = -\sn(q|k)\dn(q|k) =
%-\sqrt{(1 - \\cn^2(q|k))(k'^2 + k^2\cdot \cn^2(q|k))} = \nn \\ =
-\sqrt{(1 - H^2)(k'^2 + k^2 H^2)},
\label{cnid}
\ee
(this is because $\sn^2 q = 1 - \cn^2 q$,
$\dn^2 q = k'^2 + k^2\cn^2 q$, $k'^2 + k^2 = 1$ and
$\cn q = H$)
we get for (\ref{Calduality}):

\be
\left(\frac{\partial H}{\partial P}\right)^2 = \frac{p^2}{Q^2}
(1-H^2)(k'^2 + k^2H^2).
\ee
Now from the first eqn.(\ref{Calduality}) $p^2$ can be expressed
through $Q$ and $\sn^2(q|k) = 1 - \cn^2(q|k) = 1 - H^2$ as

\be
\frac{p^2}{Q^2} = 1 - \frac{2\nu^2}{Q^2(1- H^2)},
\ee
so that

\be
\left(\frac{\partial H}{\partial P}\right)^2
= \left( 1 - \frac{2\nu^2}{Q^2} - H^2\right)
\left( k'^2 + k^2H^2\right).
\ee
Therefore $H$ is an elliptic function of $P$, namely \cite{fgnr,bmmm3}

\be
H(P,Q) = \cn(q|k) = \alpha(Q) \cdot
\cn\left(P\sqrt{k'^2 + k^2\alpha^2(Q)}\ \bigg| \
\frac{k\alpha(Q)}{\sqrt{k'^2 + k^2\alpha^2(Q)}}\right)
\label{dualCal}
\ee
with

\be
\alpha\sp2(Q) = \alpha\sp2_{rat}(Q) = 1 - \frac{2\nu^2}{Q^2}.
\ee

In the limit $\nu^2 = 0$, when the interaction is switched off,
$\alpha(q) = 1$ and $H(P,Q)$ reduces to $H_0(P) =\ \cn(P|k)$,
as assumed in (\ref{Calduality}).

We have therefore obtained a dual formulation of the elliptic Calogero
model (in the simplest $SU(2)$ case). At first glance our dual Hamiltonian
looks somewhat unusual. In particular, the relevant elliptic curve is
``dressed'': it is described by an effective modulus

\be
k_{eff} = \frac{k\alpha(Q)}{\sqrt{k'^2 + k^2\alpha^2(Q)}}
= \frac{k\alpha(Q)}{\sqrt{1 - k^2(1- \alpha^2(Q))}},
\ee
which differs from the ``bare'' one $k$ in a $Q$-dependent way.
In fact $k_{eff}$ is nothing but the
modulus of the ``reduced''
Calogero spectral curve \cite{IM}.

Now one can rewrite (\ref{dualCal}) in many different forms, in particular,
in terms of $\theta$-functions, in hyperelliptic parametrization etc.
This latter is rather essential and we refer the reader for the details to 
the paper \cite{bmmm3}. In particular, this is the hyperelliptic 
parametrization,
where one can conveniently check that the symplectic structure $dP\wedge dQ$ 
is actually equal to $dp^{Jac}\wedge da$, i.e. to the action-angle variables
in the Calogero model. Here $p^{Jac}$, (\ref{pjac}) 
is the angle along the cycle in the Jacobian and $a$ is the action variable
given by (\ref{aad}). 

All of the above formulae are straightforwardly generalised
from the Calogero (rational-elliptic) system to the Ruijsenaars
(trigonometric-elliptic) system. The only difference ensuing is that
the $q$-dependence of the dual (elliptic-trigonometric) Hamiltonian
is now trigonometric rather than rational \cite{bmmm3}:

\be
\alpha\sp2(q) = \alpha\sp2_{trig}(q) = 1 - \frac{2\nu^2}{\sinh^2 q}
\ee
Rather than giving further details we will proceed directly to
a consideration of the double-elliptic model.

\section{Double-elliptic two-body system}

In order to get a double-elliptic system one needs to
exchange the rational $Q$-dependence
in (\ref{Calduality}) for an elliptic one, and so we substitute
$\alpha\sp2_{rat}(Q)$ by the obvious elliptic
analogue $\alpha\sp2_{ell}(Q) = 1 - \frac{2\nu^2}{\sn^2(Q|\tilde k)}$.
Moreover, now the elliptic curves for $q$ and $Q$ need not in general
be the same, i.e. $\tilde k \neq k$.

Instead of (\ref{Calduality}) the duality equations now become

\be
\cn(q|k) = H(P,Q|k,\tilde k), \nn \\
\cn(Q|\tilde k) = H(p,q|\tilde k,k), \nn \\
\cn'(Q|\tilde k)\frac{\partial H(P,Q|k,\tilde k)}{\partial P} =
\cn'(q|k)\frac{\partial H(p,q|\tilde k,k)}{\partial p},
\label{dellduality}
\ee
and the natural ansatz for the Hamiltonian (suggested
by (\ref{dualCal})) is

\be
H(p,q|\tilde k,k) = \alpha(q|\tilde k,k)\cdot
\cn\left(p\;\beta(q|\tilde k,k)\ |\ \gamma(q|\tilde k,k)\right)
= \alpha\cn(p\beta|\gamma), \nn \\
H(P,Q|k,\tilde k) = \alpha(Q|k,\tilde k)\cdot
\cn\left(P\;\beta(Q|k,\tilde k)\ |\ \gamma(Q|k,\tilde k)\right)
= \tilde\alpha \cn(P\tilde\beta|\tilde\gamma).
\label{dellans}
\ee

Substituting these ansatz into (\ref{dellduality}) and making use
of (\ref{cnid}), one can arrive, after some calculations \cite{bmmm3}, 
at\footnote{For ease of expression hereafter we suppress the dependence of
$\alpha,\beta,\gamma$ on $k$ and $\tilde k$ in what follows
using $\alpha(q)$ for $\alpha(q|\tilde k,k)$
and $\tilde\alpha(Q)$ for $\alpha(Q|k,\tilde k)$ etc.} 

\be\label{ab}
\alpha^2(q|\tilde k,k) = \alpha^2(q|k) =
1 - \frac{2\nu^2}{\sn^2(q|k)}, \ \ \ 
\beta^2(q|\tilde k,k) = \tilde k'^2 + \tilde k^2\alpha^2(q|k), \ \ \ 
\gamma^2(q|\tilde k,k) = \frac{\tilde k^2\alpha^2(q|k)}
{\tilde k'^2 + \tilde k^2\alpha^2(q|k)},
\ee
with some constant $\nu$ and finally the double-elliptic duality becomes 
\cite{bmmm3}

\be
H(P,Q|k,\tilde k)= \cn(q|k) =
\alpha(Q|\tilde k) \cn\left(
P\sqrt{k'^2 + k^2\alpha^2(Q|\tilde k)}\ \bigg|\
\frac{k\alpha(Q|\tilde k)}{\sqrt{k'^2 + k^2\alpha^2(Q|\tilde k)}}
\right),
\ee
\be\label{dellHsu2}
H(p,q|\tilde k,k) = \cn(Q|\tilde k) =
\alpha(q|k) \cn\left(p\sqrt{\tilde
k'^2 + \tilde k^2\alpha^2(q|k)}\ \bigg|\
\frac{\tilde k\alpha(q|k)}
{\sqrt{\tilde k'^2 + \tilde k^2\alpha^2(q|k)}}
\right).
\ee

We shall now consider various limiting cases arising from these
and  show that the double-elliptic Hamiltonian (\ref{dellHsu2})
contains the entire Ruijsenaars-Calogero and Toda family
as its limiting cases, as desired. (Of course we have restricted ourselves
to the $SU(2)$ members of this family.)

In order to convert the elliptic dependence of the momentum $p$
into the trigonometric one, the corresponding ``bare''
modulus $\tilde k$ should vanish: $\tilde k \rightarrow 0, \
\tilde k'^2 = 1 - \tilde k^2 \rightarrow 1$ (while $k$
can be kept finite). Then, since
$\cn(x|\tilde k = 0) = \cosh x$,

\be
H^{dell}(p,q) \longrightarrow \alpha(q)\cosh p =
H^{Ru}(p,q)
\ee
with the same

\be
\alpha^2(q|k) = 1 - \frac{2\nu^2}{\sn^2(q|k)}.
\ee
Thus we obtain the $SU(2)$ elliptic Ruijsenaars Hamiltonian.
The trigonometric and rational Ruijsenaars
as well as all of the Calogero and
Toda systems are obtained through further limiting procedures
in the standard way.

The other limit $k \rightarrow 0$ (with $\tilde k$ finite) gives
$\alpha(q|k) \rightarrow \alpha_{trig}(q) = 1 - \frac{2\nu^2}{\cosh q}$
and

\begin{equation}
\begin{array}{rl}
H^{dell}(p,q) \longrightarrow &
\alpha_{trig}(q)\cdot
\cn\left(
p\sqrt{\tilde k'^2 + \tilde k^2\alpha_{trig}^2(q)}\left|
\frac{\tilde k\alpha_{trig}(q)}
{\sqrt{\tilde k'^2 + \tilde k^2\alpha_{trig}^2(q)}}
\right.\right) = \tilde H^{Ru}(p,q).\\
\end{array}
\end{equation}
This is the elliptic-trigonometric model, dual to the
conventional elliptic Ruijsenaars (i.e. the trigonometric-elliptic)
system.
In the further limit of small $q$
this degenerates into the elliptic-rational
model with $\alpha_{trig}(q) \rightarrow \alpha_{rat}(q) =
1 - \frac{2\nu^2}{q^2}$, which is dual to the conventional elliptic
Calogero (i.e. the rational-elliptic) system, analysed
above.

Our approach has been based on choosing appropriate functions
$f(q)$ and $F(Q)$ and implementing duality. Other choices of
functions associated with alternative free Hamiltonians may be possible.
Instead of the duality relations (\ref{dellduality}) one could consider
those based on $h_0(p) = \sn(p|\tilde k)$ instead of $\cn(p|\tilde k)$.
With this choice one gets somewhat simpler expressions for $\beta_s$
and $\gamma_s$:

\be
\beta_s = 1, \ \ \
\gamma_s(q|\tilde k,k) = \tilde k\alpha_s(q|k), \ \ \
\alpha_s(q|k) = 1 - \frac{2\nu^2}{\cn^2(q|k)}
\ee
and the final Hamiltonian is now

\be
H_s(p,q|\tilde k,k) = \alpha_s(q|k) \cdot
\sn(p|\tilde k\alpha_s(q|k)).
\label{dellHamsin}
\ee
Although this Hamiltonian is somewhat simpler than our earlier choice,
the limits involved in obtaining the Ruijsenaars-Calogero-Toda
reductions are somewhat more involved, and that is why we
chose to present the Hamiltonian (\ref{dellHsu2}) first.

One might further try other elliptic functions for $h_0(p)$.
Every solution we have obtained by making a different ansatz
has been related to our solution (\ref{dellHsu2}) via modular transformations
of the four moduli $\tilde k$, $k$, $\tilde k_{eff} =
\tilde\gamma$ and $k_{eff} = \gamma$.

\section{$pq$-duality, $\tau$-functions and Hamiltonian reduction}

Now one should extend the results of the two-particle case to the generic
number of particles. Unfortunately, it is still unknown how to get very
manifest formulas. Here we describe a method that ultimately allows one to 
construct
commuting Hamiltonians of the $N$-particle Dell system, but very explicit 
expressions will be obtained only for first several terms in a perturbation
theory. 

However, in order to explain the idea, we need some preliminary work. 
In fact, we start from the old observation \cite{KriF,zeroes} that, if one
study the elliptic (trigonometric, rational) solution to the KP hierarchy,
the dynamics (dependence on higher times of the hierarchy) 
of zeroes of the corresponding $\tau$-function w.r.t. the first time
is governed by the elliptic (trigonometric, rational) Calogero system. 
Moreover, if one starts with the two-dimensional Toda lattice hierarchy
and studies the zeroes of the $\tau$-function 
w.r.t. the zero (discrete) time, the corresponding
system governing zeroes is the Ruijsenaars system \cite{KriZab}.

Let us illustrate this with the simplest example of the trigonometric solution
of the KP hierarchy. This is, in fact, the $N$-solitonic solution, the
corresponding $\tau$-function being

\be\label{soltau}
\tau(\{t_k\}|\left\{\nu_i,\mu_i,X_i\right\})=
\det_{N\times N}\left(\delta_{ij}-L_{ij}
e^{\sum_k t_k(\mu_i^k-\nu_i^k)} \right),\ \ \ 
L_{ij}\equiv {X_i\over \nu_i-\mu_j}
\ee
Here $\nu_i$, $\mu_i$ and $X_i$ are the soliton parameters. Let us impose
the constraint $\nu_i=\mu_i+\epsilon$. Then, the standard Hamiltonian 
structure 
of the hierarchy on the soliton solution gives rise to the Poisson brackets
(in \cite{FT,Babelon} similar calculation was done for the sine-Gordon case)

\be
\left\{X_i,X_j\right\}=-{4X_iX_j\epsilon^2\over (\mu_i-\mu_j)(\mu_i-\mu_j
-\epsilon)(\mu_i-\mu_j+\epsilon)},\ \ \
\left\{\mu_i,\mu_j\right\}=0,\ \ \ 
\left\{X_i,\mu_j\right\}=X_i\delta_{ij}
\ee
Note that, upon identification $\mu_i=q_i$, $X_i\equiv e^{p_i}
\prod_{l\ne i}\left({1\over (q_i-q_l)^2} -{1\over\epsilon^2}
\right)$, the matrix
$L_{ij}$ in (\ref{soltau}) becomes the rational Ruijsenaars Lax operator,
with the proper symplectic structure $\{p_i,q_j\}=\delta_{ij}$. 

At the same time, (\ref{soltau}) can be rewritten in the form

\be
\tau(\{t_k\}|\left\{\nu_i,\mu_i,X_i\right\})\sim 
\det_{N\times N}\left(e^{\epsilon x}\delta_{ij}-L_{ij} 
e^{\sum_{k\ne 1} t_k(\mu_i^k-\nu_i^k)} \right)
\ee
where $x\equiv t_1$. This determinant
is the generating function of the rational Ruijsenaars Hamiltonians. 
On the other hand, the $N$ zeroes of 
the determinant as the function of $\epsilon x$ are just logarithms of the 
eigenvalues of the Lax operator. These zeroes are governed, 
as we know from \cite{KriF,zeroes} by the trigonometric Calogero
system, and are nothing but the $N$ particle coordinates in the Calogero
system:

\be
\tau(\{t_k\}|\left\{\nu_i,\mu_i,X_i\right\})\sim 
\sum_k e^{k\epsilon x}H_k^R\sim \prod_i^N\sin ({\epsilon x}-q_i)
\ee
Therefore, the eigenvalues of the (rational Ruijsenaars) Lax operator are the 
exponentials $e^{q_i}$ of coordinates in the (trigonometric Calogero) dual 
system,
while the $\tau$-function simultaneously is the generating function of
the Hamiltonians in one integrable systems and the function of coordinates
in the dual system. This is exactly the form of duality, as it
was first realized by S.Ruijsenaars \cite{rud}. Its relation with
$\tau$-functions was first observed by D.Bernard and O.Babelon \cite{Babelon} 
(see also \cite{Khar2}).

Similarly, one can consider the $N$-solitonic solution in the two-dimensional
Toda lattice hierarchy. Then, 

\be\label{soltaut}
\tau_n(\{t_k\}|\left\{\nu_i,\mu_i,X_i\right\})=
\det_{N\times N}\left(\delta_{ij}-L_{ij}
\left({\mu_i\over\nu_i}\right)^n
e^{\sum_k t_k(\mu_i^k-\nu_i^k)+\bar t_k(\mu_i^{-k}-\nu_i^{-k})}\right),\ \ \ 
L_{ij}\equiv {X_i\over \nu_i-\mu_j}
\ee
This time one should make a reduction $\nu_i=e^{\epsilon}\mu_i$. 
Then, $L_{ij}$ becomes
the Lax operator of the trigonometric Ruijsenaars, and the zeroes w.r.t. to
$\epsilon n$ are governed by the same, trigonometric Ruijsenaars system. 
This is another check that this system is really self-dual. 

Thus constructed duality can be also interpreted
in terms of the
Hamiltonian reduction \cite{ruijh,ruo,fgnr}. In doing so, one starts 
with the moment map which typically is a constraint for two matrices or
two matrix-valued functions. In order to solve it,
one diagonalizes one of the matrices, its diagonal elements ultimately
being functions (exponentials) of the coordinates, while the other matrix
gives the Lax operator (its traces are the Hamiltonians). 
Now what one needs in order to construct the dual
system is just to diagonalize the second matrix in order to get 
coordinates in the dual system, while the first matrix will provide one with 
the new 
Lax operator (its traces are the Hamiltonians in the dual system).

As an illustrative example, let us consider the trigonometric 
Calogero-Sutherland 
system \cite{oper}. 
In order to get it, one can start from the free Hamiltonian system
$H=\tr A^2$ given on the phase space $(A,B)$ of two $N\times N$ matrices, with
the symplectic form $\tr (\delta A\wedge B^{-1}\delta B$ (i.e.
$A$ lies in the $gl(N)$ algebra, while $B$ lies in the $GL(N)$ group). 
Then, one
can impose the constraint

\be\label{mm}
A-BAB^{-1}=\nu ({\bf I}-P)
\ee
where $\nu$ is a constant, ${\bf I}$ is the unit matrix and 
$P$ is a matrix of rank 1. Now one has to make the Hamiltonian reduction 
with this
constraint.
First of all, one solves the constraint diagonalizing matrix $B$ and using
the gauge freedom of the system to transform $P$ into the matrix with all 
entries unit. It gives\footnote{Since (\ref{mm}) does not fix the
diagonal elements of $A$, they are parametrized by arbitrary numbers $p_i$.} 
$A_{ij}=p_i\delta_{ij}+(1-\delta_{ij}){\nu b_j\over b_i-b_j}$, where $b_i$ are
the diagonal elements of $B$.
This is nothing but the Lax operator of the trigonometric Calogero-Sutherland 
model.
Traces of powers of this Lax operators give the Calogero Hamiltonians.

One can also diagonalize the other matrix, $A$. Then, (\ref{mm}) can be 
rewritten
in the form

\be\label{mm2}
AB-BA=\nu (B-\bar P)
\ee
where $\bar P$ is another rank 1 matrix which is can be gauged out 
to the matrix with all entries depending only on the number of row (let denote
them
$X_i/\nu$). Solving then this constraint, one obtains
$B={X_i\over a_i-a_j-\nu}$, where $a_i$ are the diagonal elements of the 
matrix $A$.
This is nothing but the Lax operator of the rational \RS system, traces of its
powers
give a set of the dual Hamiltonians.

Certainly, this picture nicely explaining the Ruijsenaars observation
of duality is just the $\tau$-function approach told in different words.

\section{Dual Hamiltonians for many-body systems}

Unfortunately, the scheme described in the previous section does not
work for elliptic models with arbitrary number of particles. The reason is 
that even having the dual Lax operator calculated, one is typically not
able in elliptic cases to construct the dual Hamiltonians, i.e. invariant
combinations involving the Lax operator.

Instead, in order to obtain dual Hamiltonians, here 
we use the other approach discussed above, that dealing with 
zeroes of the $\tau$-functions \cite{bmmm3,MM1}. This approach gives no
very explicit form of the dual Hamiltonians. Therefore, we
construct later a kind of perturbative procedure for the Hamiltonians
which allows one to get them absolutely explicitly term by term \cite{MM1}.

Consider $SU(N)$ ($N=g+1$) system.
The whole construction is based on the fact that
the spectral curve of the original integrable system
(Toda chain, Calogero, Ruijsenaars or the most interesting double
elliptic system) has a period matrix $T_{ij}(\vec a)$,
$i,j = 1,\ldots,N$ with the special property:

\be
\sum_{j=1}^N T_{ij}(a) = \tau, \ \ \forall i
\label{sumT}
\ee
where $\tau$ does not depend on $a$.
As a corollary, the genus-$N$ theta-function is naturally decomposed
into a linear combination of genus-$g$ theta-functions:

\be
\Theta^{(N)}(p_i|T_{ij}) =
\sum_{n_i \in Z} \exp\left(i\pi\sum_{i,j=1}^NT_{ij}n_in_j +
2\pi i\sum_{i=1}^N n_ip_i\right) 
=\sum_{k=0}^{N-1}
\theta_{\left[{k\over N},{0}\right]}(N\zeta|N\tau)
\cdot \check\Theta_k^{(g)}(\check p_i|\check T_{ij})
\ee
where
\be
\Theta_k \equiv
\check\Theta_k^{(g)}(\check p_i|\check T_{ij}) \equiv
\sum_{{n_i \in Z}\atop{\sum n_i = k}}
\exp\left(-i\pi\sum_{i,j=1}^N\check T_{ij}n_{ij}^2 +
2\pi i\sum_{i=1}^N n_i\check p_i\right)
\ee
and $p_i = \zeta + \check p_i$, $\sum_{i=1}^N \check p_i = 0$;
$T_{ij} = \check T_{ij} + \frac{\tau}{N}$,
$\sum_{j=1}^N \check T_{ij} = 0$, $\forall i$.

Now we again use the argument that zeroes of the KP (Toda)
$\tau$-function (i.e. essentially the Riemannian theta-function), associated
with the spectral curve (\ref{sc}) are nothing but the coordinates $q_i$ of
the original (Calogero, Ruijsenaars) integrable system.
In more detail, due to the property (\ref{sumT}),
$\Theta^{(N)}(p|T)$ as a function
of $\zeta = \frac{1}{N} \sum_{i=1}^N p_i$ is an elliptic function on the
torus $(1,\tau)$ and, therefore, can be decomposed into an $N$-fold product
of the genus-one theta-functions. Remarkably, their arguments are just
$\zeta-q_i$:

\be
\Theta^{(N)}(p|T) =
c(p,T,\tau)\prod_{i=1}^N \theta(\zeta - q_i(p,T)|\tau)
\ee
(In the case of the Toda chain when $\tau\to
i\infty$ this ``sum rule" is implied by the standard expression
for the individual $e^{q_i}$ through the KP $\tau$-function.) Since one
can prove that $q_i$ form a Poisson-commuting set of variables with respect to
the symplectic structure (\ref{ss}) \cite{KriF}, 
this observation indirectly justifies
the claim \cite{bmmm3,MM1} that all the ratios
$\Theta_k/\Theta_l$ (in order to cancel the non-elliptic factor $c(p,T,\tau)$)
are Poisson-commuting with respect to the
Seiberg-Witten symplectic structure

\be
\sum_{i=1}^N dp_i^{Jac}\wedge da_i
\label{ss}
\ee
(where $a_N \equiv -a_1 - \ldots - a_{N-1}$, i.e.
$\sum_{i=1}^N a_i = 0$)\footnote{It can be equivalently written as
$$
\Theta_i\{\Theta_j, \Theta_k\} +
\Theta_j\{\Theta_k, \Theta_i\} +
\Theta_k\{\Theta_i, \Theta_j\} = 0  \ \ \forall i,j,k
$$
or
$$
\{\log\Theta_i, \log\Theta_j\} =
\left\{\log\frac{\Theta_i}{\Theta_j}, \log\Theta_k\right\}
\ \ \forall i,j,k
$$
}:

\be
\left\{\frac{\Theta_k}{\Theta_l}, \frac{\Theta_m}{\Theta_n}\right\} = 0
\ \ \forall k,l,m,n
\label{comm}
\ee
The Hamiltonians of the dual integrable system can be chosen in
the form $H_k = \Theta_k/\Theta_0$, $k = 1,\ldots, g$.
However, these dual Hamiltonians are not quite manifest,
since they depend on the period matrix of the original system
expressed in terms of its action variables.
Nevertheless, what is important, one can work with the $\theta$-functions 
as with
series and, using the instantonic expansion for the period matrix
$T_{ij}$, construct Hamiltonians term by term in the series.

\section{Dual Hamiltonians perturbatively}

Let us see how it really works. We start with
the simplest case of the Seiberg-Witten family, the periodic Toda chain.

\subsection{Perturbative approximation}

First we consider the perturbative approximation which is nothing but the 
open Toda chain, see (\ref{pertcurv}) and (\ref{pertdS}).
This is the case of the perturbative $4d$ pure $N=2$ SYM theory with
the prepotential

\be
{\cal F}_{pert}(a) = \frac{1}{2i\pi}\sum_{i<j}^N
a_{ij}^2\log a_{ij}
\ee
Then, the period matrix
is singular and only finite number of terms survives in the
series for the theta-function:

\be
\Theta^{(N)}(p|T) = \sum_{k=0}^{N-1} e^{2\pi ik\zeta}H^{(0)}_k(p,a),
\label{pertTodatheta}
\ee
\be
H^{(0)}_k(p,a) = \sum_{I, [I]=k} \prod_{i\in I}e^{2\pi ip_i}
\prod_{j\in \bar I} {\cal Z}^{(0)}_{ij}(a)
\label{pertTodaHam}
\ee
Here
\be
{\cal Z}^{(0)}_{ij}(a)=e^{-i\pi T^{(0)}_{ij}} = \frac{\Lambda_{QCD}}{a_{ij}}
\label{pertTodaF}
\ee
and $I$ are all possible partitions of $N$ indices into the sets
of $k = [I]$ and $N-k = [{\bar I}]$ elements. Parameter $\Lambda_{QCD}$
becomes significant only when the system is deformed:
either non-perturbatively or to more complex systems of the
Calogero--Ruijsenaars--double-elliptic  family.
The corresponding $\tau$-function $\Theta^{(N)}(p|T)$,
eq.(\ref{pertTodatheta}), describes an $N$-soliton solution
to the KP hierarchy and is equal to the determinant (\ref{soltau}). 
The Hamiltonians $H^{(0)}_k$,
eq.(\ref{pertTodaHam}) dual to the open Toda chain, 
are those of the degenerated rational Ruijsenaars
system, and they are well-known to Poisson-commute with respect
to the relevant Seiberg-Witten symplectic structure (\ref{ss}).
This is not surprising since the open Toda chain is obtained by a 
degeneration from
the trigonometric Calogero system which is dual to the rational Ruijsenaars.

The same construction for other perturbative
Seiberg-Witten systems ends up with the Hamiltonians
of more sophisticated systems.

For the trigonometric Calogero system
(perturbative $4d$ $\CN=4$ SYM with SUSY softly broken down
to $\CN=2$ by the adjoint mass $M$) the Poisson-commuting (with respect to 
the same
(\ref{ss})), whose perturbative prepotential is given by the sum of 
(\ref{pppg}) and (\ref{pppa}), Hamiltonians $H^{(0)}_k$ are given by
(\ref{pertTodaHam}) with

\be
{\cal Z}^{(0)}_{ij}(a) = \frac{\sqrt{a_{ij}^2 - M^2}}{a_{ij}}
\label{pertCalF}
\ee
i.e. are the Hamiltonians of the rational Ruijsenaars system which are really
dual to the trigonometric Calogero model.

For the period matrix (effective charges) (\ref{3.67}) 
of the trigonometric Ruijsenaars system
(perturbative $5d$ $\CN=2$ SYM compactified on a circle
with an $\epsilon$ twist as the boundary conditions)
the Hamiltonians are given by (\ref{pertTodaHam}) with

\be
{\cal Z}^{(0)}_{ij}(a) =
\frac{\sqrt{\sinh(a_{ij}+\epsilon)\sinh(a_{ij}-\epsilon)}}
{\sinh a_{ij}}
\label{pertRuF}
\ee
i.e. are the Hamiltonians of the trigonometric Ruijsenaars system, which
is, indeed, self-dual. 

Finally, for the perturbative limit of the
most interesting self-dual double-elliptic system
(the explicit form of its spectral curves, whose period matrix is supposed 
to be
(\ref{3.68}), is yet unknown but in
the case of two particles, see the next section)
the relevant Hamiltonians are those of the elliptic Ruijsenaars
system, given by the same (\ref{pertTodaHam}) with (cf. with (\ref{3.68}))

\be
{\cal Z}^{(0)}_{ij}(a) =\sqrt{1-{2g^2\over \hbox{sn}^2_{\tilde\tau}(a_{ij})}}
\sim\frac{\sqrt{\theta(\hat a_{ij}+\varepsilon|\tilde\tau)
\theta(\hat a_{ij}-\varepsilon|\tilde\tau)}}
{\theta(\hat a_{ij}|\tilde\tau)}
\label{pertdellF}
\ee
where $\tilde\tau$ is the modulus of the second torus associated with the
double elliptic system and $\hat a_{ij}$ are $a_{ij}$ rescaled, \cite{BEWW}. 

In all these examples $H_0^{(0)} = 1$, the theta-functions
$\Theta^{(N)}$ are singular and given by determinant (solitonic)
formulas with finite number of items (only terms with $n_i =0, 1$
survive in the series expansion of the theta-function),
and Poisson-commutativity of arising Hamiltonians is analytically
checked within the theory of Ruijsenaars integrable systems.

\subsection{Beyond the perturbative limit}

Beyond the perturbative limit, the analytical evaluation of
$\Theta^{(N)}$ becomes less straightforward.

The non-perturbative deformation (\ref{fsc-Toda}) of the curve 
(\ref{pertcurv}),
is associated with somewhat sophisticated prepotential
of the periodic Toda chain,

\be
{\cal F}(a) = \frac{1}{4\pi i}\sum_{i<j}^N a_{ij}^2\log\frac{a_{ij}}{\Lambda}
+ \sum_{k=1}^\infty \Lambda^2{\cal F}^{(k)}(a)
\ee
The period matrix is

\be
\pi iT_{ij} = \frac{\partial^2{\cal F}}{\partial a_i\partial a_j} =
-2\pi i\tau\delta_{ij}\ + \log\frac{a_{ij}}{\Lambda} +
\sum_{k=1}^\infty \Lambda^2
\frac{\partial^2{\cal F}^{(k)}(a)}{\partial a_i\partial a_j},
\ \  i\neq j, \\ T_{ii} =-2 \tau - \sum_{j\neq i} T_{ij}
\ee
and $\tau$ in this case can be removed by the rescaling of $\Lambda$.
Then,

\be
\Theta^{(N)}(p|T) = \sum_{k=0}^{N-1} e^{2\pi ik\zeta}
\Theta_k(p,a) 
= \sum_{k=0}^{N-1} e^{2\pi i k\zeta}
\sum_{n_i,\ \sum_i n_i = k}
e^{-i\pi\sum_{i<j} T_{ij}n_{ij}^2} e^{2\pi i \sum_i n_ip_i} = \\
= \left( 1 + \sum_{i\neq j}^N e^{2\pi i(p_i - p_j)}
{\cal Z}_{ij}^4\prod_{k\neq i,j} {\cal Z}_{ik}{\cal Z}_{jk} + \right. \\
\left. +
\sum_{i\neq j\neq k\neq l}^N  e^{2\pi i(p_i + p_j - p_k - p_l)}
{\cal Z}_{ik}^4 {\cal Z}_{il}^4 {\cal Z}_{jk}^4  {\cal Z}_{jl}^4
\prod_{m\neq i,j,k,l} {\cal Z}_{im}{\cal Z}_{jm}{\cal Z}_{km}{\cal Z}_{lm}
+ \ldots\ \right) + \\
+ e^{2\pi i \zeta}
\left(\sum_{i=1}^N e^{2\pi i p_i} \prod_{j\neq i} {\cal Z}_{ij} +
\sum_{i\neq j\neq k} e^{2\pi i (p_i + p_j - p_k)}
{\cal Z}_{ik}^4{\cal Z}_{jk}^4\prod_{l\neq i,j,k} {\cal Z}_{il}{\cal Z}_{jl}
{\cal Z}_{kl} + \right.\\
\left. + \sum_{i\neq j} e^{2\pi i (2p_i - p_j)}
{\cal Z}_{ij}^9\prod_{k\neq i,j} {\cal Z}_{ik}^4 {\cal Z}_{jk} + \ldots \
\right) + \\
+ e^{4\pi i\zeta}
\left(\sum_{i\neq j} e^{2\pi i(p_i+p_j)}\prod_{k\neq i,j}{\cal Z}_{ik}
{\cal Z}_{jk} +
\sum_i e^{4\pi ip_i} \prod_{k\neq i}{\cal Z}_{ik}^4 + \ldots \ \right) + \\
+ \ldots
\label{Thetaexp}
\ee
with

\be
{\cal Z}_{ij} = e^{-i\pi T_{ij}} = {\cal Z}^{(0)}_{ij}
\left( 1 -
i\pi\frac{\partial^2{\cal Z}^{(1)}(a)}{\partial a_i\partial a_j}
+ \ldots\ \right), \ \ \      i\neq j
\label{F}
\ee
The first few corrections  ${\cal Z}^{(k)}$ to the prepotential
are explicitly known in the Toda-chain case \cite{IC},
for example,

\be
{\cal Z}^{(1)} = -\frac{1}{2i\pi}\sum_{i=1}^N\prod_{k\neq i}
\left({\cal Z}_{ik}^{(0)}\right)^2
\label{F1}
\ee

The coefficients $\Theta_k$ in (\ref{Thetaexp})
are expanded into powers of $(\Lambda/a)^{2N}$ and the leading (zeroth-order)
terms are exactly the perturbative expressions (\ref{pertTodaHam}).
Thus, the degenerated Ruijsenaars Hamiltonians (\ref{pertTodaHam}) are just
the
perturbative approximations to the $H_k = \Theta_k/\Theta_0$ --
the full Hamiltonians of the integrable system dual to the Toda
chain.

The same expansion can be constructed for the other, elliptic systems.
Note that the period matrix in these cases is expanded into powers of
$q=e^{2\pi i\tau}$:

\be
\pi iT_{ij} = \frac{\partial^2{\cal F}}{\partial a_i\partial a_j} =
-2\pi i\tau\delta_{ij}\ + \log\frac{a_{ij}}{\Lambda} +\mu^2
\sum_{k=1}^\infty q^k
\frac{\partial^2{\cal F}^{(k)}(a)}{\partial a_i\partial a_j},
\ \  i\neq j
\ee
and the dimensional constant $\mu$ depends on the system. E.g., in the
Calogero
model $\mu=iM$ etc. The limit to the Toda system corresponds to $q\to 0$, $q
\mu^{2N}$=fixed.

Note that the first (instanton-gas) correction to the prepotential
is indeed known to be of the same universal form (\ref{F1}) not only for the 
Toda
chain, but also  for the Calogero system \cite{DPhongCP}. For the
Ruijsenaars and double-elliptic systems eq.(\ref{F1})
is not yet available in the literature. 

In \cite{MM1} the perturbative expansions for the Hamiltonians were
tested for Poisson commutativity, up to few first terms and small number of 
particles $N$. In particular, it was checked that
(\ref{F1}) is true for both the Ruijsenaars and Dell systems.

The main lesson we could get from this consideration is that
if the {\it universal} expressions like (\ref{F1})
in terms of perturbative ${\cal Z}^{(0)}_{ij}$, the same   for all
the systems, will be found for higher corrections
to the prepotentials\footnote{
To avoid possible confusion, the recurrent relation
\cite{LNS,RG} for the Toda-chain prepotential,
$$
\frac{\partial^2{\cal F}}{\partial\log\Lambda^2} \sim
\frac{\partial^2{\cal F}}{\partial\log\Lambda\partial a_i}
\frac{\partial^2{\cal F}}{\partial\log\Lambda\partial a_j}
\left.\frac{\partial^2}{\partial p_i\partial p_j}
\log\Theta_0\right|_{p=0}
$$
does {\it not} immediately provide such  {\it universal}
expressions. Already for ${\cal Z}^{(1)}$ this relation gives
$$
{\cal Z}^{(1)} \sim\sum_{i<j}^N \left({\cal Z}^{(0)}_{ij}\right)^2
\prod_{k\neq i,j}^N  {\cal Z}^{(0)}_{ik}{\cal Z}^{(0)}_{jk}
$$
which coincides with (\ref{F1}) for the Toda-chain
${\cal Z}^{(0)}_{ij}$, eq.(\ref{pertTodaF}), but is not true
(in variance with (\ref{F1})) for Calogero ${\cal Z}^{(0)}_{ij}$,
eq.(\ref{pertCalF}). Meanwhile, the recurrent relations of ref.\cite{MNW} for
the Calogero system provide more promising expansion.
},  this will immediately give an explicit
(although not the most appealing) construction of the
Hamiltonians dual to the Calogero and Ruijsenaars models and, especially
important, the self-dual double-elliptic system which
was explicitly constructed above only for $N=2$.

Let us note now that the 
$SU(2)$ Hamiltonian (\ref{dellHsu2}) has a rather nice form in terms
of the $p,q$ variables or their duals $P,Q$.
However, this our general $SU(N)$ construction of the Hamiltonians as ratios
of
genus $g$ $\theta$-functions uses another kind of
canonical variables -- angle-action variables $p^{Jac}_i, a_i$.
The {\it flat} moduli $a_i$ play the central role in 
\SW theory, while $Q_i$ are rather generalizations
of the algebraic moduli.

It is an interesting open problem to express the Hamiltonians
in terms of $P_i$ and $Q_i$, perhaps, they can acquire a more
transparent form, like it happens for $SU(2)$.
Another option is to switch to the
``separated" variables ${\cal P}_i$, ${\cal Q}_i$ \cite{sklyanin} such that
$\oint_{A_j}{\cal P}_id{\cal Q}_i=a_i\delta_{ij}$ (while generically
$a_i=\oint_{A_j}\sum_i P_idQ_i$). Generically, they are different from $P_i$,
$Q_i$ but coincide with $P$, $Q$ in the case of $SU(2)$.

\section{Dell systems: 6d adjoint matter}

Thus, we have constructed the Dell system and a set of dualities that
acts on the Toda-Calogero-Ruijsenaars-Dell family. They are all
put together in Fig.2.

\begin{figure}[t]
%TexCad Options
%\grade{\on}
%\emlines{\on}
%\beziermacro{\off}
%\reduce{\on}
%\snapping{\off}
%\quality{2.00}
%\graddiff{0.01}
%\snapasp{1}
%\zoom{1.00}
\special{em:linewidth 0.4pt}
\unitlength 1mm
\linethickness{0.4pt}
\begin{picture}(141.33,120.33)
\emline{18.33}{110.00}{1}{141.33}{110.00}{2}
\emline{141.33}{110.00}{3}{141.33}{110.00}{4}
\emline{18.33}{95.00}{5}{141.33}{95.00}{6}
\emline{18.33}{80.00}{7}{141.33}{80.00}{8}
\emline{18.33}{65.00}{9}{141.00}{65.00}{10}
\emline{141.00}{120.00}{11}{141.00}{65.00}{12}
\emline{110.33}{120.00}{13}{110.33}{65.00}{14}
\emline{80.00}{120.33}{15}{80.00}{65.00}{16}
\emline{50.00}{120.00}{17}{50.00}{65.00}{18}
\emline{50.00}{110.00}{19}{18.33}{120.00}{20}
%\vector(70.67,91.67)(84.00,101.33)
\put(84.00,101.33){\vector(4,3){0.2}}
\emline{70.67}{91.67}{21}{84.00}{101.33}{22}
%\end
%\vector(85.67,99.33)(73.33,90.33)
\put(73.33,90.33){\vector(-4,-3){0.2}}
\emline{85.67}{99.33}{23}{73.33}{90.33}{24}
%\end
%\vector(102.67,76.33)(114.00,85.33)
\put(114.00,85.33){\vector(4,3){0.2}}
\emline{102.67}{76.33}{25}{114.00}{85.33}{26}
%\end
%\vector(115.33,83.33)(105.67,75.67)
\put(105.67,75.67){\vector(-4,-3){0.2}}
\emline{115.33}{83.33}{27}{105.67}{75.67}{28}
%\end
%\bezvec{96}(54.67,102.67)(47.67,112.00)(58.33,105.33)
\put(58.33,105.33){\vector(2,-1){0.2}}
\emline{54.67}{102.67}{29}{53.40}{104.44}{30}
\emline{53.40}{104.44}{31}{52.52}{105.86}{32}
\emline{52.52}{105.86}{33}{52.02}{106.94}{34}
\emline{52.02}{106.94}{35}{51.90}{107.67}{36}
\emline{51.90}{107.67}{37}{52.17}{108.05}{38}
\emline{52.17}{108.05}{39}{52.82}{108.08}{40}
\emline{52.82}{108.08}{41}{53.85}{107.77}{42}
\emline{53.85}{107.77}{43}{55.27}{107.11}{44}
\emline{55.27}{107.11}{45}{58.33}{105.33}{46}
%\end
%\bezvec{108}(84.33,87.67)(76.33,97.67)(88.67,91.33)
\put(88.67,91.33){\vector(2,-1){0.2}}
\emline{84.33}{87.67}{47}{83.03}{89.38}{48}
\emline{83.03}{89.38}{49}{82.07}{90.81}{50}
\emline{82.07}{90.81}{51}{81.46}{91.96}{52}
\emline{81.46}{91.96}{53}{81.20}{92.83}{54}
\emline{81.20}{92.83}{55}{81.28}{93.43}{56}
\emline{81.28}{93.43}{57}{81.72}{93.74}{58}
\emline{81.72}{93.74}{59}{82.50}{93.77}{60}
\emline{82.50}{93.77}{61}{83.64}{93.52}{62}
\emline{83.64}{93.52}{63}{85.12}{92.99}{64}
\emline{85.12}{92.99}{65}{88.67}{91.33}{66}
%\end
%\bezvec{80}(114.00,73.33)(108.33,81.00)(117.33,76.67)
\put(117.33,76.67){\vector(2,-1){0.2}}
\emline{114.00}{73.33}{67}{112.81}{75.06}{68}
\emline{112.81}{75.06}{69}{112.08}{76.42}{70}
\emline{112.08}{76.42}{71}{111.81}{77.40}{72}
\emline{111.81}{77.40}{73}{112.00}{78.00}{74}
\emline{112.00}{78.00}{75}{112.65}{78.23}{76}
\emline{112.65}{78.23}{77}{113.75}{78.08}{78}
\emline{113.75}{78.08}{79}{115.31}{77.56}{80}
\emline{115.31}{77.56}{81}{117.33}{76.67}{82}
%\end
%\bezvec{224}(74.00,76.67)(108.00,77.67)(114.67,98.67)
\put(114.67,98.67){\vector(1,4){0.2}}
\emline{74.00}{76.67}{83}{77.17}{76.81}{84}
\emline{77.17}{76.81}{85}{80.21}{77.04}{86}
\emline{80.21}{77.04}{87}{83.13}{77.36}{88}
\emline{83.13}{77.36}{89}{85.92}{77.77}{90}
\emline{85.92}{77.77}{91}{88.60}{78.27}{92}
\emline{88.60}{78.27}{93}{91.15}{78.86}{94}
\emline{91.15}{78.86}{95}{93.57}{79.54}{96}
\emline{93.57}{79.54}{97}{95.88}{80.31}{98}
\emline{95.88}{80.31}{99}{98.06}{81.17}{100}
\emline{98.06}{81.17}{101}{100.11}{82.12}{102}
\emline{100.11}{82.12}{103}{102.05}{83.16}{104}
\emline{102.05}{83.16}{105}{103.86}{84.29}{106}
\emline{103.86}{84.29}{107}{105.55}{85.51}{108}
\emline{105.55}{85.51}{109}{107.11}{86.82}{110}
\emline{107.11}{86.82}{111}{108.55}{88.22}{112}
\emline{108.55}{88.22}{113}{109.87}{89.72}{114}
\emline{109.87}{89.72}{115}{111.07}{91.30}{116}
\emline{111.07}{91.30}{117}{112.14}{92.97}{118}
\emline{112.14}{92.97}{119}{113.09}{94.73}{120}
\emline{113.09}{94.73}{121}{113.92}{96.58}{122}
\emline{113.92}{96.58}{123}{114.67}{98.67}{124}
%\end
%\bezvec{220}(117.00,96.67)(107.67,75.00)(76.00,73.33)
\put(76.00,73.33){\vector(-1,0){0.2}}
\emline{117.00}{96.67}{125}{116.05}{94.64}{126}
\emline{116.05}{94.64}{127}{115.01}{92.70}{128}
\emline{115.01}{92.70}{129}{113.86}{90.85}{130}
\emline{113.86}{90.85}{131}{112.60}{89.10}{132}
\emline{112.60}{89.10}{133}{111.25}{87.44}{134}
\emline{111.25}{87.44}{135}{109.79}{85.87}{136}
\emline{109.79}{85.87}{137}{108.23}{84.39}{138}
\emline{108.23}{84.39}{139}{106.57}{83.00}{140}
\emline{106.57}{83.00}{141}{104.81}{81.71}{142}
\emline{104.81}{81.71}{143}{102.94}{80.50}{144}
\emline{102.94}{80.50}{145}{100.97}{79.39}{146}
\emline{100.97}{79.39}{147}{98.90}{78.37}{148}
\emline{98.90}{78.37}{149}{96.72}{77.44}{150}
\emline{96.72}{77.44}{151}{94.45}{76.60}{152}
\emline{94.45}{76.60}{153}{92.07}{75.86}{154}
\emline{92.07}{75.86}{155}{89.58}{75.21}{156}
\emline{89.58}{75.21}{157}{87.00}{74.64}{158}
\emline{87.00}{74.64}{159}{84.31}{74.17}{160}
\emline{84.31}{74.17}{161}{81.53}{73.80}{162}
\emline{81.53}{73.80}{163}{78.63}{73.51}{164}
\emline{78.63}{73.51}{165}{76.00}{73.33}{166}
%\end
\put(39.33,119.00){\makebox(0,0)[cc]{{coordinate}}}
\put(26.67,113.33){\makebox(0,0)[cc]{{momentum}}}
\put(64.67,114.33){\makebox(0,0)[cc]{rational}}
\put(95.00,113.67){\makebox(0,0)[cc]{trigonometric}}
\put(125.00,114.00){\makebox(0,0)[cc]{elliptic}}
\put(33.67,102.00){\makebox(0,0)[cc]{rational}}
\put(33.33,86.67){\makebox(0,0)[cc]{trigonometric}}
\put(33.33,72.67){\makebox(0,0)[cc]{elliptic}}
\put(73.00,102.00){\makebox(0,0)[cc]{\parbox{.27\linewidth}{rational
Calogero}}}
\put(98.67,102.00){\makebox(0,0)[cc]{\parbox{.19\linewidth}
{trigonometric Calogero}}}
\put(130.67,102.33){\makebox(0,0)[cc]{\parbox{.15\linewidth}{elliptic
Calogero}}}
\put(65.67,86.67){\makebox(0,0)[cc]{\parbox{.2\linewidth}{rational
Ruijsenaars}}}
\put(95.67,86.33){\makebox(0,0)[cc]{\parbox{.15\linewidth}
{trigonometric Ruijsenaars}}}
\put(129.67,86.33){\makebox(0,0)[cc]{\parbox{.18\linewidth}
{elliptic Ruijsenaars}}}
\put(64.67,72.67){\makebox(0,0)[cc]{\parbox{.17\linewidth}
{dual Calogero}}}
\put(98.67,71.00){\makebox(0,0)[cc]{\parbox{.15\linewidth}
{dual Ruijsenaars}}}
\put(129.67,73.00){\makebox(0,0)[cc]{\parbox{.2\linewidth}{Dell system}}}
\end{picture}
\vspace{-6.8cm}
\caption{Action of the coordinate-momentum duality on the
Calogero-Ruijsenaars-Dell family. Hooked arrows mark self-dual
systems. The duality leaves the coupling constant $\nu$ intact.}\label{dual}
\end{figure}
\vspace{10pt}

Since the theories of this family describes low energy limits of different 
SUSY gauge theories, this table describes the dualities between different
gauge theories too. Say, perturbative limit of the $4d$ theory (trigonometric 
Calogero) is dual to a special degeneration \cite{bm}
of the perturbative limit of the $5d$ theory (rational Ruijsenaars), while 
full, {\it non-perturbative} $5d$ theory (elliptic Ruijsenaars) is dual to
the perturbative limit of the $6d$ theory (Dell system).

To conclude our discussion of dualities and Dell systems, we calculate the
perturbative prepotential of the 2-particle Dell system (\ref{dellHsu2})
and show that it coincides with what is to be expected for the $6d$ theory
with adjoint matter, supporting the identification of the Dell system with the
$6d$ gauge theory.

Let us start from the full Dell model for 2-particles. Then, it is
given by the Hamiltonian (\ref{dellHsu2})
parametrized by two independent (momentum and coordinate) 
elliptic curves with elliptic moduli $k$ and $\tilde k$.  
As usual, for two particles we can construct the {\it full}
spectral curve from the Hamiltonian (\cite{GKMMM}, see also section 4 and
(\ref{times})),

\be\label{sc}
H(\zeta,\xi|k,\tilde k) = u \ \ \ \ \left(\ = \cn(Q|k)\ \right)
\ee
It is characterized by the effective elliptic moduli

\be
k_{eff} = \frac{k \alpha(q|\tilde k)} {\beta(q|k,\tilde k)},\ \ \ \
\tilde k_{eff} = \frac{\tilde k \alpha(q|k)}
{\beta(q|\tilde k,k)}
\ee
where the functions $\alpha$ and $\beta$ are manifestly given in (\ref{ab}).
Coordinate-momentum duality interchanges
$k \leftrightarrow \tilde k$, $k_{eff} \leftrightarrow \tilde k_{eff}$
(and $q,p \leftrightarrow Q,P$), while
in general $SU(N)$ case, the model describes an interplay between
the four tori: the two bare elliptic curves and two
effective Jacobians of complex dimension $g = N-1$.

The generating differential in $6d$ theories is of the form
$dS=\zeta d\xi$, with $\xi$ living on the coordinate torus and
$\zeta$ -- on the momentum torus.

Now  we calculate the leading order of the prepotential expansion in
powers of $\tilde k$ when the bare spectral torus degenerates into
sphere. In the
forthcoming calculation we closely follow the line of \cite{bmmm2}.

When $\tilde k\to 0$, $\sn(q|\tilde k)$ degenerates into the ordinary sine.
For further convenience, we shall
parametrize the coupling constant $2\nu^2\equiv\sn^2(\epsilon|k)$. Now the
spectral curve (\ref{sc}) acquires the form

\be\label{psc}
\alpha(\xi)\equiv\sqrt{1-{\sn^2(\epsilon|k)\over\sin^2\xi}}=
{u\over \cn
\left(\zeta\beta\bigg|k_{eff}\right)}
\ee
Here the variable $\xi$ lives in the cylinder produced after degenerating
the bare coordinate torus. So does the variable $x=1/\sin^2\xi$. Note
that the $A$-period of the dressed torus shrinks on the sphere to a contour
around $x=0$. Similarly, $B$-period can be taken as a contour passing from
$x=0$ to $x=1$ and back.

The next step is to calculate variation of the generating differential
$dS=\eta d\xi$ w.r.t. the modulus
$u$ in order to obtain a holomorphic differential:

\be
dv=\left(-i\sn(\epsilon|k)\sqrt{k'^2+k^2u^2}\right)^{-1}
{dx\over x\sqrt{(x-1)(U^2-x)}}
\ee
where $U^2\equiv{1-u^2\over\sn^2(\epsilon|k)}$.
Since

\be
{\partial a\over\partial u}=\oint_{x=0}{\partial dS\over\partial u}
=\oint_{x=0}dv=
-{1\over \sqrt{(1-u^2)(k'^2+k^2u^2)}}
\ee
we deduce that $u=\cn (a|k)$ and $U={\sn(a|k)\over\sn(\epsilon|k)}$. The
ratio of the $B$- and $A$-periods of $dv$
gives the period matrix

\be
T={U\over\pi}\int_0^1 {dx\over x\sqrt{(x-1)(U^2-x)}}=
-{1\over i\pi}\lim_{\kappa\to 0}\left(\log{\kappa\over 4}\right)+
{1\over i\pi}\log{U^2\over 1-U^2}
\ee
where $\kappa$ is a small-$x$ cut-off. The $U$ dependent part of this
integral is finite and can be
considered as the \lq\lq true" perturbative correction, while
the divergent part just renormalizes the bare  \lq\lq classical"
coupling constant $\tau$, {\it i.e.} classical part of the prepotential
(see \cite{bmmm2} for further details).
Therefore, the perturbative period matrix is finally

\be
T_{finite}={i\over\pi}\log{\sn^2(\epsilon|k)-
\sn^2 (a|k)\over\sn^2 (a|k)}+\ const\
\longrightarrow
{i\over\pi}\log{\theta_1(a+\epsilon)\theta_1(a-\epsilon)\over\theta_1^2(a)}
\ee
and the perturbative prepotential is the elliptic tri-logarithm.
Remarkably, it lives on
the {\it bare} momentum torus, while the modulus of the perturbative curve
(\ref{psc}) is the dressed one.

\paragraph{Acknowledgements.}

The author is grateful to H.W.Braden, A.Gorsky, S.Gukov, A.Marshakov, 
A.Morozov, I.Krichever for discussions and collaboration 
and to S.Kharchev, M.Olshanetsky and A.Zabrodin for useful
discussions. I am also obliged to T.Takebe for kind hospitality at the
Ochanomizu University, Tokyo, where this work was completed. 
The work was supported in part by grants INTAS 99-0590, CRDF \#6531, 
RFBR 00-02-16101a and the JSPS fellowship for research in Japan.

\newpage
\renewcommand{\refname}{References}

\end{document}